%% file: cbncbv.tex
\RequirePackage{fix-cm}
\makeatletter
\def\cl@chapter{}
\makeatother
\documentclass[smallcondensed]{svjour3}     % onecolumn (ditto)
\smartqed  % flush right qed marks, e.g. at end of proof
\usepackage{graphicx}
\usepackage{amsmath,amssymb}
\usepackage{stmaryrd}
\usepackage{tikz}
\usetikzlibrary{shapes,calc}
\usepackage[hyphens]{url}
\usepackage{multirow}
\usepackage{multicol}
\usepackage{rotating}
\usepackage{listings}
\usepackage{lmodern} % Set all fonts to Latin Modern
\usepackage{mathptmx} % Reset text and math fonts to Times (load AFTER lmodern)
\usepackage[utf8]{inputenc}
\usepackage[T1]{fontenc}
\DeclareSymbolFont{letters}{OML}{txmi}{m}{it}
\DeclareMathAlphabet{\mathcal}{OMS}{cmsy}{m}{n}
\usepackage{bussproofs}
\usepackage{hyphenat}
\usepackage{enumitem} 
\usepackage{fancyvrb}
\usepackage{listings}

\makeatletter
\g@addto@macro \normalsize {%
	\setlength\abovedisplayskip{2.2pt}%
	\setlength\belowdisplayskip{2.2pt}%
}

\newenvironment{myitem}[1][]
{\itemize[leftmargin=2.5ex,topsep=0.3ex,itemsep=1pt, #1]}
{\enditemize}

\makeatother

\let\llncssubparagraph\subparagraph
\let\subparagraph\paragraph
\usepackage[compact]{titlesec}
\let\subparagraph\llncssubparagraph
\usepackage{titlesec}
\titlespacing*{\section}{0.5pt}{1.1\baselineskip}{0.3\baselineskip}
\titlespacing*{\subsection}{0.5pt}{0.65\baselineskip}{0.25\baselineskip}

\usepackage{color}

\definecolor{light-gray}{gray}{0.85}
\newcommand\coll[1]{\mbox{\colorbox{light-gray}{$\!#1\!$}}}

\usepackage{amsthm} 

\usepackage{varioref}

\usepackage[
a4paper,
pdftex,  
pdfkeywords={},
pdfborder={0 0 0},
draft=false,
bookmarksnumbered,
bookmarks,
bookmarksdepth=2,
bookmarksopenlevel=2,
bookmarksopen]{hyperref}

\usepackage{cleveref}
\usepackage{cite}
\usepackage{calc}

\include{myCommands}

\allowdisplaybreaks

\allowdisplaybreaks
\usepackage{thmtools}
\usepackage{thm-restate}
\usepackage{enumitem}

\makeatletter
\def\thm@space@setup{%
	\thm@preskip=1.22ex
	\thm@postskip=\thm@preskip % or whatever, if you don't want them to be equal
}
\makeatother

\begin{document}
	
	\title{
		Case Studies in Formal Reasoning About Lambda-Calculus: 
		%Syntax with Bindings 
		\\Semantics, Church-Rosser, Standardization and HOAS
		%A Case Study in Reasoning about Syntax with Bindings: 
	%	\\The %$\lambda$-Calculus 
	%	Church-Rosser and  Standardization Theorems
	}
	
	\titlerunning{Case Studies in Reasoning about }        % if too long for running head
	
	\author{Lorenzo Gheri         \and
		Andrei Popescu %etc.
	}

	%Institute of Mathematics Simion Stoilow of the Romanian Academy, Bucharest, Romania

	%\authorrunning{Short form of author list} % if too long for running head
	
	\institute{Lorenzo Gheri \at
		%first address \\
		%Tel.: +123-45-678910\\
		%Fax: +123-45-678910\\
		Department of Computing
		\\
		Imperial College London
		\\\email{l.gheri@imperial.ac.uk}           %  \\
		%             \emph{Present address:} of F. Author  %  if needed
		\and
		Andrei Popescu \at
		Department of Computer Science
		\\
		University of Sheffield
		\\\email{a.popescu@sheffield.ac.uk}
	}
	
	\date{ 
		%Received: date / Accepted: date
	}
	% The correct dates will be entered by the editor

	\maketitle
%%%%%%%%%%%%%

\begin{abstract}
%\vspace*{-5ex}
We have previously published the Isabelle/HOL formalization of a general theory of syntax with bindings. 
In this companion paper, we instantiate the general theory to the syntax of lambda-calculus and formalize the development leading to   
several fundamental constructions and results: sound semantic interpretation, the Church-Rosser and standardization 
theorems, and higher-order abstract syntax %(HOAS) 
 encoding. For Church-Rosser and standardization, our work covers both the call-by-name and call-by-value versions of the calculus, following 
classic papers by Takahashi and Plotkin. 
During the formalization, we were able to stay focused on the high-level 
ideas of the development---thanks to the arsenal provided by %for free 
our general theory:  %essential use of the recursion principles and Nominal-style 
a wealth of basic facts about the substitution, swapping and freshness operators, 
as well as recursive-definition and reasoning principles, including a specialization to semantic interpretation of syntax.  
\end{abstract}

%\vspace*{-5ex}
\section{Introduction}

Formal reasoning about syntax with bindings is a notoriously challenging problem, due to 
the difficulty of handling binding-specific aspects such as alpha-equivalence (also known 
as naming equivalence), capture-avoiding substitution of terms for variables, 
and the generation of variables that are fresh in certain contexts. 

Informal techniques 
aimed at easing the reasoning tasks have turned out to be very difficult 
to represent formally, partly due to their reliance on unstated assumptions without which they 
would be unsound. For example, the majority of textbooks on $\lambda$-calculi (including the most standard one \cite{bar-lam}) employ the principle of 
primitive recursion to define functions on $\lambda$-terms, after which they tacitly assume 
these functions to be invariant under alpha-equivalence; as another example, the so-called 
Barendregt variable convention assumes that, in a proof or definition context, 
the bound variables are fresh for all the parameters located outside the scope of their binders. 
Both these principles are unsound in general, that is, if employed without checking some sanity conditions on  
the defining clauses or on the definition and proof context.  

Formal reasoning frameworks 
have been designed to recover such informal principles %techniques 
on a sound basis. The approaches range 
from a clever manipulation of the bound variables 
as in nominal logic and the locally named representation 
\cite{pitts01nominal,urban-NominalHOL,pol-LocNamed2} to 
the removal of the very notion of bound variable---by either encoding away bound variables as numeric positions in terms as in de Bruijn-style and locally nameless representations 
%positionwise as in de Bruijn-style and locally nameless %/named techniques 
\cite{bru-lam,fio-abs,locallyNamelessOverview} 
%\cite{locallyNameless} 
or by representing them using meta-variables as in higher-order abstract syntax (HOAS) 
\cite{phe-hig,har-fra,pau-genTh,DBLP:conf/cade/PfenningS99,weakHOAS,momFelty-Hybrid4,chlipala-Parametric,feltyPientka-comparison}.   

Our own framework \cite{ghepop-2017-jar} takes a  nominal-style approach.
%, but differs from Nominal Logic in a important aspects, including its operator-aware recursion principles and the presence of built-in substitution.  
%
The framework is formalized in the Isabelle/HOL proof assistant as a many-sorted theory parameterized over a binding signature. 
Its distinguishing features 
(some of which also set it apart from nominal logic) 
are a rich built-in theory of 
substitution, swapping 
and freshness, as well as recursion and semantic interpretation principles that are sensitive to these operators. 

In previous work, we have deployed our framework to formalize classic results in many-sorted first-order logic (completeness of deduction and soundness of 
Skolemization  \cite{blanchette-et-al-2014-ijcar,soundCompl-jou,blanchette-frocos2013})  
and System F (strong normalization \cite{pop-HOASOnFOAS}), and novel results about   
the meta-theory of Isabelle's Sledgehammer 
tool \cite{blanchette-frocos2013,blanchette-et-al-2013-types}. However, in the papers describing these applications we have emphasized neither 
(1) the general theory underlying our framework nor (2) the framework's deployment to support reasoning  within these applications.  
The first gap has been filled in a recent paper \cite{ghepop-2017-jar}. The second gap is being filled by the current paper, which is intended as a companion to \cite{ghepop-2017-jar}. 
  
This paper presents the instantiation of the  framework to support the development of some fundamental constructions and results in $\lambda$-calculus with $\beta$-reduction: soundness of semantic interpretation, the Church-Rosser and 
standardization theorems, 
and adequacy of a HOAS encoding.\footnote{We emphasize that this is a case study in formalizing the meta-theory of HOAS-style encoding; our framework itself does not follow the HOAS methodology.} 
The Church-Rosser and 
standardization theorems are established for both the call-by-name and call-value variants.\footnote{Our formalization is publicly available from
the paper's website~\cite{lambda-scripts}.}  

The first step we take is instantiating the framework to the syntaxes of  
call-by-name and call-by-value $\lambda$-calculus, the latter differing from 
the former by the existence of an additional syntactic category of special terms called values.  
%as a particular kind of terms. 
 These instantiations provide us with a rich theory of the standard operators on terms, namely freshness, substitution and swapping, as well as a  freshness-aware induction proof principle and  operator-aware recursive definition principles, including a variant specialized to semantic interpretation (Section~\ref{sec-inst}).  

Then we proceed with the formal development of our specific target results. 
We only show in detail the development for the call-by-name calculus (Section~\ref{sec-CBN}). 
The similar Church-Rosser and standardization development for the call-by-value calculus is only sketched by pointing out the differences, including the use of a two-sorted instantiation of our framework  (Section~\ref{sec-CBV}). 
 
The results require the definition of 
standard $\beta$-reduction and $\beta$-equivalence (Section~\ref{sub-CBN-Reds}), including variations such as parallel and left $\beta$-reduction.
%, single-step and multi-step reductions. 
%
Semantic interpretation is defined in Henkin-style models, and takes full advantage of our framework's built-in semantic features (Section~\ref{sub-CBN-sound}). 
The Church-Rosser theorem (Section~\ref{sub-CBN-CR}) is proved by formalizing the parallel-reduction technique of Tait \cite{bar-lam}, enhanced with the complete parallel reduction operator trick due to  Takahashi \cite{takahashi-CompleteDevelopment}. 
For standardization (Section~\ref{sub-CBN-Std}), we follow closely Plotkin's original paper \cite{plotkin-CBNandCBVandLambda}. 
As HOAS case study, we consider a simple encoding of $\lambda$-calculus in itself (Section~\ref{sub-CBN-HOAS}).  

Our presentation emphasizes the use of the various principles provided by our framework, as well as some difficulties arising from representing formally some informal definition and proof idioms---such as recursing over alpha-equated terms (or, equivalently, recursing in an alpha-equivalence preserving manner) and inversion rules obeying Barendregt's variable convention. Some of the lessons learned during the formalization effort, as well as some statistics, are presented in Section~\ref{sec-overview}. 
We conclude with an overview of related work  
%, including a more detailed %focused 
%comparison with the very related work of Isabelle's %Nominal package 
(Section~\ref{sec-relWork}).

\section{Instantiation of the General Framework}
\label{sec-inst}

Our framework \cite{ghepop-2017-jar} is parameterized by a binding signature, which 
essentially specifies the following data: a collection of term sorts, a collection of variable 
sorts,\footnote{Even though variables of all sorts behave essentially the same, they are delivered as different collections, belonging to different sorts. For example, this allows one to sharply distinguish between individual and set variables in second-order logic, or between channel names and process names in process calculi.} an embedding relationship between variable sorts and term sorts, 
and a collection of (term) constructors, each with an assigned arity and an assigned result sorts.

%The arity consists of zero or more input sorts---where an input sort is either 
%just a term sort (for free inputs) or a pair of a variable sort and a term sort (for bound 
%inputs). The result sort refers to the output of the constructor. 

%The signature parameter are fixed in an Isabelle locale. 
The theory was  
developed over an arbitrary signature, which is represented as an Isabelle locale \cite{Locales}. 
Namely, ``quasi-terms'' were defined as being freely generated by the constructors, then terms were
defined by quotienting quasi-terms to the notion of alpha-equivalence obtained 
standardly from the signature-specified bindings of the term constructors. 
Thus, what we call ``terms'' in this paper are alpha-equivalence classes.  
Several standard operators were defined on terms, including capture-avoiding substitution 
of terms for variables, freshness of a variable for a term, and swapping of two variables 
in a term. The theory provides many properties of these operators, as well as binding-aware 
and standard-operator-aware structural 
recursion and induction principles and a principle for interpreting 
syntax in a semantic domain. 

Our companion paper \cite{ghepop-2017-jar} gives details about this general framework. However, 
understanding %the framework
these details is not necessary for following the rest of this paper, which gives 
a %fairly 
self-contained description of two instances of the framework. 
%In fact, the description of these instances will give 
%the reader a good idea about the capabilities of our framework in general. 

\subsection{The syntax of $\lambda$-calculus} \label{sub-inst-CBN}

Our first instance is the paradigmatic syntax of $\lambda$-calculus (with constants), which is 
typically informally specified 
using a grammar such as  
\[
\begin{array}{rcl}
X    &\;::=\;& \Var\;x \;\mid\; \Ct\;c \;\mid\; \App\;X\;Y  \;\mid\; \Lm\;x\;X 
\end{array} 
\]
where $X$ and $Y$ range over terms (the ones generated by the grammar), $x$ over 
a given infinite type $\var$ of variables and $c$ over a given type $\const$ of constants---where $\Var$ and $\Ct$ are the embeddings of variables and constants into 
terms, $\App$ is application and $\Lm$ is $\lambda$-abstraction. 
Terms are assumed to be 
equated modulo alpha-equivalence, defined standardly by assuming that, in $\Lm\;x\;X$, 
the $\lambda$-constructor $\Lm$ binds the variable $x$ in the term $X$. Thus, for example, 
$\Lm\;x\;(\Var\;x) = \Lm\;y\;(\Var\;y)$ even if $x \not=y$. 

We obtain the above syntax by 
picking a particular binding signature (with a single sort of variables and a single sort of terms, and, with the desired constructors). 
In Isabelle, picking a signature corresponds to instantiating the corresponding locale. 
In addition to this straightforward instantiation, 
we also perform a formal transfer of all the concepts and results to a more shallow (and hence more usable) Isabelle representation. This involves creating native Isabelle/HOL types 
of terms for each sort of the signature and transferring all the term constructors and operators and all facts about them to these native types. The process is conceptually straightforward, but is quite tedious, and must be done by hand since we have not yet automated it. 
\cite[\S6.5]{ghepop-2017-jar} offers more details, 
and \cite[\S5]{schropp-nonfree} presents the automation of a similar kind of transfer (for nonfree datatypes). 

For our instance of interest ($\lambda$-calculus with constants), this results 
in the type $\term$ of $\lambda$-terms together with: 
\begin{myitem}
	\item the constructors, namely $\Var : \var \ra \term$, 
	 	$\Ct : \const \ra \term$,  
	 	$\App : \term \ra \term \ra \term$ 
	 	and $\Lm : \var \ra \term \ra \term$ 
 \item and the standard operators:
 \begin{myitem}
 	\item depth (height) of a term, $\depthh : \term \ra \nat$ 
 	\item freshness of a variable in a term,\footnote{Other frameworks employ a free-variable operator, $\textsf{FVars} : \term \ra \var\;\sett$. This is of course inter-definable with the freshness operator.}  
 	$\fresh : \var \ra \term \ra \bool$
 	%unary 
 	\item (capture-avoiding) substitution of a term for 
 	a variable in a term, $\_[\_ / \_] : \term \ra \term \ra \var \ra \term$
 	\item (capture-avoiding) parallel substitution of multiple terms for multiple variables in a term, $\_[\_] : \term \ra (\var \ra \term\;\option) \ra \term$
 	\item swapping of two variables in a term,\footnote{While not explicitly present in the traditional $\lambda$-calculus \cite{bar-lam}, swapping has been popularized by nominal logic as a very convenient operator in bootstrapping definitions---thanks to the fact that 
 		bijective renamings behave better than arbitrary renamings with respect to 
 		bindings \cite{pitts-AlphaStructural}.} $\_[\_ \swap \_] : \term \ra \var \ra \var \ra \term$
\end{myitem}
\end{myitem}

\vspace*{0.5ex}
From our general theory, we also obtain for free: 
\begin{myitem}
	\item many basic facts proved about the constructors and operators
	\item and induction and recursion principles 
	for proving new facts about terms and defining new functions on terms, respectively  
\end{myitem}

Our framework provides %all the imaginable 
a multitude of general-purpose properties of the constructors and operators, including properties about their mutual 
interactions.  
%\footnote{We challenge the reader to find one useful property about these operators that is not accounted for %provided 
%	in our framework \cite{lambda-scripts}.} 
%
For example, the following are two essential properties of equality between $\lambda$-abstractions, reflecting the fact that 
terms are alpha-equivalence classes. The second 
%of them 
allows us to rename bound variables with fresh ones, whenever needed.   

\begin{prop}  \label{prop-CBN-quasi-inj}\ 
The following hold: 
\\(1) If $y \notin \{x,x'\}$ and $\fresh\;y\;X$ 
and 
 $\fresh\;y\;X'$ and $ X\,[(\Var\;y)\,/\,x] = X'\,[(\Var\;y)\,/\,x']$ then 
$\Lm\;x\;X = \Lm\;x'\;X'$ 
\\(2) If $\fresh\;y\;X$ then $\Lm\;x\;X = \Lm\;y\;(X\,[(\Var\;y)\,/\,x])$. 
\end{prop}

Another example is the compositionality of substitution: 

\begin{prop}  \label{prop-CBN-subst_comp}\ 
The following hold: 
\\(1) %Substitution of the same variable distributes over itself: 
%
%$$
$X\ [Y_1 \,/\, y]\, [Y_2 \,/\, y] \;=\; X\,[(Y_1\,[Y_2 \,/\, y]) / y]$
%$$
%
%\noindent
\\(2)
%(2) Substitution of different variables distributes over itself, assuming %suitable 
%freshness: % conditions:
%
%$$
If $y \neq z$ and $\fresh\;y\;Z$ then 
$X\,[Y \,/\, y]\, [Z \,/\, z] \,=\, X\,[Z \,/\, z]\,[(Y\,[Z \,/\, z]) \,/\, y]$
%$$
\end{prop}

\vspace*{0.5ex}
\noindent
{\bf Fresh structural induction.} Our framework also offers a structural induction principle in the style of nominal logic \cite{pitts-AlphaStructural,UrbanTasson,urban-Barendregt}. It differs from standard structural induction  
in that, in the inductive $\Lm$-case, it allows one to additionally assume  
freshness of the $\Lm$-bound variable with respect to any potential parameters of the to-be-proved statement. For the $\lambda$-calculus instance, it becomes:

\begin{prop}\label{th-CBN-fresh-ind} 
	{\rm (Fresh structural induction principle)}
	Let $\param$ be a type (of items called parameters) endowed with a function $\varsOf : \param \ra \var\;\sett$ such that $\varsOf\;p$ is finite for all $p:\param$. 
	Let $\phi : \term \ra \param \ra \bool$ be a predicate on terms and parameters.  
		
	Assume the following four sentences are true for all $x:\var$, $c:\const$ and  
	$X,Y : \term$:
	\\(1) $\phi\;(\Var\;x)\;p$ holds for all $p:\param$. 
	\\(2) $\phi\;(\Ct\;c)\;p$ holds for all $p:\param$.
	\\(3) If $\phi\;X\;p$ and $\phi\;Y\;p$ hold for all $p:\param$, 
	then $\phi\,(\App\;X\;Y)\;q$ holds for all 
	$q:\param$. 
	\\(4) If $\phi\;X\;p$ holds for all $p:\param$, 
	then $\phi\,(\Lm\;x\;X)\;q$ holds for all 
	$q:\param$ such that 
	$\coll{x \not\in \varsOf\;q}$. 
	
	Then $\phi\;X\;p$ holds for all $X:\term$ and $p:\param$. 
\end{prop}

For details on the wide applicability of this parameter-based fresh induction principle we refer the reader to 
\cite{UrbanTasson}. The parameters are typically taken to be the other terms and variables appearing in a statement, different from the term on which we induct. 
A classic example 
%, discussed in \cite[page 2]{UrbanTasson}, 
is the proof of substitution compositionality, our 
Prop.~\ref{prop-CBN-subst_comp}(2)---which can be done by fresh induction on $X$ taking as parameters all the other terms and variables, namely $Y,y,Z$ and $z$. In the $\Lm$-case, thanks to the extra freshness assumption, we can soundly invoke Barendregt's variable convention and assume, for example, that in the expression $(\Lm\;x\;X)\,[Y \,/ \,y]\, [Z \,/\, z]$ we have $x$ fresh for $Y,y,Z$ and $z$---which allows reducing the expression to  $\Lm\;x\;(X\;[Y \,/\, y]\, [Z \,/\, z])$ and then applying the induction hypothesis.  By contrast, applying standard induction would have brought serious complications concerning variable renaming.  

Prop.~\ref{th-CBN-fresh-ind}  
immediately implies the following fresh 
case distinction principle. It states that any term is either a variable, or a constant, or an application, or an abstraction whose bound variable can be taken to be fresh for a given parameter.   

\begin{prop}\label{th-CBN-fresh-case} 
	{\rm (Fresh case distinction principle)}
	Let $\param$ and $\varsOf$ be like in the 
	previous proposition and let $Z:\term$ and $p:\param$. 
	Then one of the following holds:  
	\\(1) $Z = \Var\;x$ for some  $x:\var$. 
	\\(2) $Z = \Ct\;c$ for some $c:\const$.
	\\(3) $Z = \App\;X\;Y$ for some $X,Y:\term$.
	\\(4) $Z = \Lm\;x\;X$ for some $x:\var$ and $X:\term$ such that $\coll{x \not\in \varsOf\;p}$.
\end{prop}

%When employing the above principles, the type $\param$ is often instantiated to $(\var + \term)\;\llist$, in order to capture any number of 
%variable and/or term parameters in the to-be-proved statements. 

\vspace*{1ex}
\noindent
{\bf Operator-aware recursion.} 
Our framework offers structural recursion principles for defining functions $H$ from terms 
to any other target type, based on the following ingredients:
\begin{myitem}
	\item a description of the recursive behavior of $H$ with respect to the term 
	constructors (as is common with primitive recursion on free datatypes)
	\item a description of the expected interaction of $H$ with freshness on the one hand and 
	substitution and/or swapping on the other hand
\end{myitem}
These are achieved by organizing the target type as a ``model'' that interprets 
the constructors and the operators in specific ways.

\newcommand\fake[1]{#1}
\begin{defi} \label{def-FSb}
A {\em freshness-substitution model (FSb model)} is a type $\DDD$ 
endowed with the following: 
 \begin{myitem}
 	\item functions on $\DDD$ having similar types 
 	as the term constructors (but with $\term$ replaced with $\DDD$ in their 
 	target type and with the pair of $\term$ and $\DDD$ in their source types), 
 	namely $\VAR : \var \ra \DDD$, 
 	$\CT : \const \ra \DDD$,  
 	$\APP : \term \ra \DDD \ra \term \ra \DDD \ra \DDD$ 
 	and $\Lm : \var \ra \term \ra \DDD \ra \DDD$ 
 	\item functions on $\DDD$ having similar types 
 	as the freshness and substitution operators (again, with $\term$ suitably 
 	replaced with $\DDD$ or with $\term$ and $\DDD$), namely 
 	$\FRESH : \var \ra \term \ra \DDD \ra \bool$ 
 	and $\SSUBST : \term \ra \DDD \ra \term \ra \DDD \ra \var \ra \DDD$ 
 \end{myitem}
The above functions are allowed to be defined in any way, provided they satisfy 
the following freshness clauses (F1)-(F5), substitution clauses (Sb1)--(Sb4) 
and substitution-renaming clause (SbRn): 
\begin{description}
	\item{F1:} $\FRESH\ x\ \fake{(\Ct\ c)}\ (\CT\ c)$
	\item{F2:} $x\not=z$ implies $\FRESH\;z\;\fake{(\Var\;x)}\;(\VAR\;x)$ 
	\item{F3:} $\FRESH\;z\;\fake{X'}\;X$ and $\FRESH\;z\;\fake{Y'}\;Y$ implies $\FRESH\;z\;\fake{(\App\;X'\;Y')}\;(\APP\;\fake{X'}\;X\;\fake{Y'}\;Y)$
	\item{F4:} $\FRESH\;z\;\fake{(\Lm\;z\;X')}\;(\LM\;z\;\fake{X'}\;X)$ 
	\item{F5:} $\FRESH\;z\;\fake{X'}\;X$ implies $\FRESH\;z\;\fake{(\Lm\;x\;X')}\;(\LM\;x\;\fake{X'}\;X)$
	\item{Sb1:} $\SSUBST\ (\Var\;z)\ (\VAR\;z)\ Z'\ Z\ z = Z$
	\item{Sb2:} $x\not=z$ implies 
	$\SSUBST\ (\Var\;x)\ (\VAR\;x)\ Z'\ Z\ z = \VAR\;x$
	\item{Sb3:} $\SSUBST\ (\App\;X'\;Y')\ (\APP\;X'\;X\;Y'\;Y)\ Z'\ Z\ z =$\ \\
	$\APP\;(X'[Z'\ /\ z])\;(\SSUBST\ X'\ X\ Z'\ Z\ z)\;(Y'[Z'\ /\ z])\;(\SSUBST\ Y'\ Y\ Z'\ Z\ z)$
	\item{Sb4:} $x \not=z$ and $\FRESH\;x\;Z'\;Z$ implies 
	\\$\SSUBST\ (\Lm\;z\;X')\ (\LM\;x\;X'\;X)\ Z'\ Z\ z = \LM\;x\;(X'[Z'\ /\ z])\;(\SSUBST\ X'\ X\ Z'\ Z\ z)$
	%%%%
	\item{SbRn:} $x\not=y$ and $\FRESH\;y\;X'\;X$ implies 
	\\$\LM\;y\;(X'[(\Var\;y)\, / x])\;(\SSUBST\ X'\ X\ (\Var\;y)\ (\VAR\;y)\ x) = \LM\;x\;X'\;X$
\end{description} 
\end{defi}

\begin{defi} \label{def-Fsw}
	A {\em freshness-swapping model (FSw model)} is similar to 
	an FSb model, except that it has a swapping-like function
	$\SSWAP : \term \ra \DDD \ra \var \ra \var \ra \DDD$ instead of the substitution-like function 
	$\SSUBST$ and satisfies the following swapping clauses (Sw1)--(Sw4) 
	and swapping-congruence clause (SwCg) instead of the substitution-related clauses 
	(Sb1)--(Sb4) and (SbRn):
\begin{description}
	\item{Sw1:} $\SSWAP\;\fake{(\Ct\ c)}\;(\CT\ c)\;z_1\;z_2\;=\;\CT\ c$  
	\item{Sw2:} $\SSWAP\;\fake{(\Var\;x)}\;(\VAR\;x)\;z_1\;z_2\;=\; \VAR\;(x\,[z_1 \swap z_2])$
	\item{Sw3:} $\SSWAP\;\fake{(\App\;X'\;Y')}\;(\APP\;\fake{X'}\;X\;\fake{Y'}\;Y)\;z_1\;z_2\ =$\ \\
	$\APP\;
	\fake{(X'\,[z_1 \swap z_2])}\,
	(\SSWAP\;\fake{X'}\;X\;z_1\;z_2)\,
	\fake{(Y'\,[z_1 \swap z_2])}\,
	(\SSWAP\;\fake{Y'}\;Y\;z_1\;z_2)$ 
	\item{Sw4:} $\SSWAP\;\fake{(\Lm\;x\;X')}\;(\LM\;x\;\fake{X'}\;X)\;z_1\;z_2\,=\,
	\LM\;(x\,[z_1 \swap z_2])\ \fake{(X'\,[z_1 \swap z_2])}\ (\SSWAP\;\fake{X'}\;X\;z_1\;z_2)$
	\item{SwCg:} 
	$\FRESH\;z\;\fake{X'}\;X$ and 
	$\FRESH\;z\;\fake{Y'}\;Y$ and 
	$z \notin \{x,y\}$ and 
	$\SSWAP\;\fake{X'}\;X\;z\;x = \SSWAP\;\fake{Y'}\;Y\;z\;y$
	implies $\LM\;x\;\fake{X'}\;X = \LM\;y\;\fake{Y'}\;Y$
	%\item{SwRn:} 
	%$\FRESH\;y\;\fake{X'}\;X$ implies $\LM\;x\;\fake{X'}\;X = \LM\;y\;\fake{(X'\,[y \swap x])}\;(\SSWAP\;y\;x\;\fake{X'}\;X)$
\end{description}
\end{defi}

To simplify notation, in what follows we will often refer to FSb models and FSw models simply by their carriers and leave the additional structure implicit, thus writing, e.g., ``Let $\DDD$ be an FSb model.''  
The framework's recursion principles essentially say that terms form the 
initial FSb and FSw models:\footnote{The reason why we define our models' operations 
to act not only on the models' carrier type $\DDD$ but also on $\term$ 
is to achieve  
the higher flexibility of {\em primitive recursion} compared to 
{\em iteration}---see \cite[\S 1.4.2]{pop-thesis} 
for a detailed discussion of this distinction. 
}  

\begin{prop}\label{prop-CBN-rec}
Let $\DDD$ be an FSb model (FSw model, respectively).
Then there exists a unique function $H : \term \ra \DDD$ commuting with the constructors, i.e.,
\begin{myitem}
  \item $H\;(\Var\ x) = \VAR\;x$
  \item $H\;(\Ct\ c) = \CT\;c$ 
  \item $H\;(\App\;X\;Y) = \APP\;X\;(H\;X)\;Y\;(H\;Y)$ 
  \item $H\;(\Lm\;x\;X) = \LM\;x\;X\;(H\;X)$ 
\end{myitem}
Additionally, $H$ preserves freshness 
and commutes with substitution (respectively, swapping):
\begin{myitem}
	\item $\fresh\;x\;X$ implies $\FRESH\;x\;X\;(H\;X)$ 
	\item $\ H\;(X[Z\,/\,z]) = \SSUBST\;X\;(H\;X)\;Z\;(H\;Z)\;z$ 
	\\(respectively, $H\;(X[z_1\swap z_2]) = \SSWAP\;X\;(H\;X)\;z_1\;z_2$)  
\end{myitem}
\end{prop}

The principle is much easier to use in practice than 
%it was to formulate:  
its elaborate formulation might suggest: 
Say one wishes to define a function $H$ from $\term$ to a type $\DDD$. Then the 
functions on $\DDD$ corresponding to the term constructors can be determined 
from the desired recursive clauses for $H$. Moreover, the functions on $\DDD$ 
corresponding to freshness and substitution or swapping are determined by 
the desired behavior of $H$ with respect to these operators, obtained from 
answering questions such as ``How can $H\,(X[Z\ /\ x])$ be expressed in terms of 
$H\;X$, $H\;Z$ and $x$?''. 

We illustrate this methodology by a simple example. (More explanations and examples can be found in \cite{pop-recPrin} 
and \cite{ghepop-2017-jar}, and in this paper's Section~\ref{sub-CBN-CR}.)
Namely, we define $\textsf{\rm no} : \term \ra \var \ra \nat$, where 
$\textsf{\rm no}\;X\;x$ counts the number of (free) occurrences of the variable $x$ in the term $X$. We do this using our recursion principle: 

\begin{defi} \label{no-def}
	$\no : \term \ra (\var \ra \nat)$ is the unique function 
	satisfying the following properties:  
	\[
	\begin{array}{l}
	\no\ (\Var\ y)\ x= \begin{cases}
	1,  \mbox{ \rm \ if \ } x=y
	\\
	0,  \mbox{ \rm \ if \ } x\not=y
	\end{cases}
	\ \ \ \ \ \ \ \ 
	\no\ (\Ct\ c)\ x =0
	\\
	\no\ (\App\ X\ Y)\ x= \no\ X\ x + \no\ Y\ x 
	\ \ \ \ \ \ \ \
	\no\ (\Lm\ y\ X)\ x = \begin{cases}
	0,  \mbox{ \rm \ if \ } x=y
	\\
	\no\ X\ x,  \mbox{ \rm \ if \ } x\not=y
	\end{cases}
	\\
	\fresh\ x\ X\ \mbox{ \rm implies } \no\ X\ x = 0
	\ \ \ \ \ \ \ \
	\no\ (X[Y \,/\, y])\ x = 
	\begin{cases}
	\no\ X\ y \,*\, \no\ Y\ y, \mbox{ \rm \ if \ } x=y
	\\
	\no\ X\ x \,+\, \no\ X\ y \,*\, \no\ Y\ x, \mbox{ \rm \ if \ } x\not=y
	\end{cases} 
	\end{array}
	\]
\end{defi}

Before formally justifying this definition (i.e., proving that there exists a unique function $\no$ satisfying the above clauses), let us explain how the clauses have been produced. First, the clauses for the constructors ($\Var$, $\Ct$, $\App$ and $\Lm$) are simply describing the desired recursive behavior of $\no$---which would have been the same had the terms not been considered modulo alpha-equivalence, but as a datatype freely generated from these constructors. However, the problem here is that the terms {\em are} quotiented, so the constructor clauses are not {\em a priori} guaranteed to form a correct definition. This is where the remaining clauses, for freshness and substitution, come into play. 
They have been produced by answering to the following questions: {\em If} the operator $\no$ was already defined, how would it behave w.r.t.\ freshness and substitution? More precisely:
\begin{myitem}
\item What would $\fresh\;x\;X$ imply about the value of $\no\;X$? Answer: It would imply that 
this value is $0$ at $x$. 
\item 
What would the value of $\no\,(X[Y \,/\, y])$ be, expressed in terms of $\no\,X$, $\no\,Y$ and $y$? 
Answer: For each variable $x$, the formula 
depends on whether $x$ is equal to $y$, and is the one shown in Def.~\ref{no-def}. (This can be easily discovered by drawing a picture of a presumptive term $X$ and the free occurrences of $y$ in it, all of which are to be substituted by $Y$.) 
\end{myitem}

In short, performing a recursive definition in our framework requires:
\begin{myitem}
	\item 
a routine part, providing the clauses for the constructors, which are immediate if one knows what one wants to define, and 
\item a somewhat creative (although often easy) ``anticipatory'' part, describing the behavior of the desired operator w.r.t.\ freshness and substitution or swapping
\end{myitem} 

To formally justify the above definition, we extract 
an FSb model obtained from the above clauses in a completely routine fashion. Namely, we take $\D = \var \ra \nat$, and  
define
$\VAR : \Var \ra \D$ and $\SSUBST : \term \ra \DDD \ra \term \ra \DDD \ra \var \ra \DDD$ by  
$$
\begin{array}{cc}
\VAR\ y\ x= \begin{cases}
	1,  \mbox{ \rm \ if \ } x=y
	\\
	0,  \mbox{ \rm \ if \ } x\not=y
	\end{cases}
&
\hspace*{6ex}
\SSUBST\;X\;u\;Y\;v\;y = \lambda x.\; 
%\no\ (X[Y\ /\ y])\ x = 
\begin{cases}
u\ y \,*\, v\ y, \mbox{ \rm \ if \ } x=y
\\
u\ x \,+\, u\ y \,*\, v\ x, \mbox{ \rm \ if \ } x\not=y
\end{cases} 
\end{array}
$$
and similarly for the other constructors and operators.

Verifying Prop.~\ref{prop-CBN-rec}'s conditions is routine---some simple arithmetics that has been discharged by Isabelle's ``auto'' proof method. 
This allows us to apply the conclusion of Prop.~\ref{prop-CBN-rec}, obtaining a unique function $\no : \term \ra \D$ commuting with the constructors, freshness and substitution---which precisely means satisfying the clauses listed in 
Def.~\ref{no-def}. 

Note again how we included {\em as part of the definition} not only the recursive clauses for the constructors, but also 
those for the interaction with freshness and substitution.  
On the one hand, the freshness and substitution clauses are needed to establish the correctness of the definition; on the other hand, they are 
useful theorems that are produced (and proved) at definition time together with the recursive clauses for the constructors.   
%These properties will turn out to be very 
%useful. 

%This methodology is explained in \cite{pop-recPrin} 
%and \cite{ghepop-2017-jar} and will be further illustrated in this paper. 

Now, let us look at some (partial) non-examples. First, consider a function $h : \term \ra \nat$ such that $h\;X$ counts the number of free variables of $X$. It can be of course immediately defined as the cardinal of $\{x \mid \neg\;\fresh\;x\;X\}$, but trying to define it recursively would be difficult (and unnatural)---since we do not have enough information to compute $h\;(\App\;X\;Y)$ from $h\;X$ and $h\;Y$. (We could ``force'' such a definition by initially counting the variable overlap between $X$ and $Y$, but this would defeat our purpose, since it would require a function more complicated than $h$.)

The above non-example applies to our recursion principle, but also to the standard recursion for free datatypes. A more subtle %(partial) 
non-example is the $\depthh$ operator, which we discuss in \cite{pop-recPrin}.\footnote{Incidentally, this operators is actually  
	built in our framework, 
	so the user has no need to define it.} This can be easily defined recursively for the free datatatype of non-quotiented terms, 
as well as for the quotiented terms
if we use the swapping-based variant of our recursion principle (with FSw-models). However, it cannot be defined using our substitution-based variant (with FSb models), since we cannot express the value of 
$\depthh\,(X[Y \,/\, y])$ from those of $\depthh\,X$ and $\depthh\,Y$; so in this case 
the problem is created not by the constructors,  
but by the substitution operator.  

\vspace*{1ex}
\noindent
{\bf Refinements of recursion.}
An advantage of our systematic, clause-based take on recursion\footnote{More precisely, what we have here are first-order theories consisting of Horn clauses \cite{pop-thesis}.} is the possibility to add  optional ``packages'' that deliver additional properties about the defined functions. 

\begin{defi}\label{def-FS-extra} 
	An FSb model (FSw model, respectively) is called {\em freshness-reversing}, if 
		it satisfies the converses of the clauses F2--F5 in Def.~\ref{def-FSb}
		 (Def.~\ref{def-Fsw}, respectively),   namely: 	
\begin{description}
	%\item{F1c:} $\FRESH\ x\ \fake{(\Ct\ c)}\ (\CT\ c)$
	\item{F2c:} 
	$\FRESH\;z\;\fake{(\Var\;x)}\;(\VAR\;x)$ implies
	$x\not=z$  
	\item{F3c:} 
	$\FRESH\;z\;\fake{(\App\;X'\;Y')}\;(\APP\;\fake{X'}\;X\;\fake{Y'}\;Y)$ 
	implies 
	$\FRESH\;z\;\fake{X'}\;X$ and $\FRESH\;z\;\fake{Y'}\;Y$ 
	\item{F4$\_$5c:} $\FRESH\;z\;\fake{(\Lm\;x\;X')}\;(\LM\;x\;\fake{X'}\;X)$ implies 
	 $x = z$ or $\FRESH\;z\;\fake{X'}\;X$ 
	\end{description} 
It is called {\em constructor-injective} if 
its constructor-like operators are injective and mutually exclusive, in that  
	\begin{myitem}
		\item  $\CT\ c$, $\VAR\ x$, $\APP\;X'\;X\;Y'\;Y$ and $\LM\;z\;Z'\;Z$ are all distinct
		\item $\CT$, $\VAR$, $\APP$ and $\LM$ are all injective (if we regard $\APP$ and $\LM$ as uncurried operators, of 4 and 3 arguments, respectively)  
	\end{myitem}	
\end{defi}

The clauses in the above definition are of course satisfied by the term model. F1c--F3c and F4$\_$5c correspond to inversion properties of freshness w.r.t.\ the constructors. Note that, being the converse of the ``direct'' clauses F4 and F5, the clause F4$\_$5c has a  disjunction as its conclusion.  

\begin{prop}\label{prop-rec-extra}
	Let $\DDD$ be an FSb model (FSw model, respectively) and let $H$ be the induced recursive function described in 
	Prop.~\ref{prop-CBN-rec}. Then the following hold: 
	\begin{myitem}
		\item If $\DDD$ is freshness-reversing, then $H$ (not only preserves, but also) reflects freshness, in that 
		$\FRESH\;x\;X\;(H\;X)$ implies 
		$\fresh\;x\;X$. 
		\item If $\DDD$ is constructor-injective, then $H$ is injective. 
	\end{myitem}
\end{prop}

The two points of Prop.~\ref{prop-rec-extra} are, just like Prop.~\ref{prop-CBN-rec}, statements of initiality properties (in different categories). This time, terms are being characterized as the initial object in:
\begin{myitem}
\item the category of freshness-reversing FSb (FSw) models and freshness-reflecting model morphisms 
\item the category of constructor-injective FSb (FSw) models and injective model morphisms 	
\end{myitem}

\vspace*{1ex}
\noindent
{\bf Interpretation in semantic domains.} 
Our general framework caters for the semantic interpretation of terms. 
A semantic domain is a structure consisting of a type for each sort and of a function for each constructor except for the variable-injection one---in such as way that binding inputs in the constructors become  second-order inputs in the associated functions.  
%For example, 
%the semantic counter-part of $\Lm : \var \ra \term \ra \term$ is an operator 
%of type $(\var \ra \SSS) \ra \SSS$, where $\SSS$ is in this case the type corresponding to the unique sort of the $\lambda$-calculus signature. 
For our particular $\lambda$-calculus syntax, this instantiates to the following concept:  

\begin{defi} \label{def-sem}
	A {\em semantic domain} is a type $\SSS$
 endowed with the 
 functions $\ct : \const \ra \SSS$, 
 $\app : \SSS \ra \SSS \ra \SSS$ and $\lm : (\SSS \ra \SSS) \ra \SSS$ 
 (corresponding to the term constructors $\Ct$, $\App$ and $\Lm$). 
\end{defi}

Just like for FSb and FSw models, we will often refer to semantic domains simply by their carriers $\SSS$, leaving the additional structure implicit. 
The following proposition allows for the interpretation of terms in any semantic domain. It was established generally, for an arbitrary syntax, by appealing to the FSb-based recursion principle. 
Here is the instance for this syntax:\footnote{In the following definition, we write  $\lambda$ for meta-level functional abstraction, and of course continue to use \textsf{Lm} for the syntactic constructor.}  

\begin{prop}\label{prop-sem} 
	Let $\SSS$ %$(\SSS,\ct,\app,\lm)$ 
	be a semantic domain, and let $\val$ be the type of valuations of %the 
	variables in the domain, 
	$\var \ra \SSS$. 
	Then 
	there exists the unique function 
	$\sem : \term \ra \val \ra \SSS$ 
	such that:
	\begin{myitem}
		\item $\sem\;(\Var\;x)\;\rho = \rho\;x$
		\item $\sem\;(\Ct\;c)\;\rho = \ct\;c$
		\item $\sem\;(\App\;X\;Y)\;\rho =
		\app\;(\sem\;X\;\rho)\;(\sem\;Y\;\rho)$ 
		\item $\sem\;(\Lm\;x\;X)\;\rho = \lm\,(\lambda s.\;\sem\;X\;(\rho[x \la s]))$  
	\end{myitem}
where $\rho[(x \la s]$ is the function $\rho$ updated at $x$ with $d$---which sends 
$x$ to $d$ and any other $y$ to $\rho\;y$.
	\par
	In addition, the interpretation 
	satisfies the following properties:  
	\begin{myitem}
		\item $\sem\;(X[Y\,/\,y])\;\rho = \sem\;X\;(\rho[y \la \sem\;Y\;\rho])$
		\item $\fresh\;x\;X$ and  $\rho=_{x}\rho'$ imply $\sem\;X\;\rho = \sem\;X\;\rho'$ 
	\end{myitem}
	where 
``$=_{x}\!$'' means ``equal everywhere except perhaps on $x$''; namely $\rho=_{x}\rho'$ holds 
iff $\rho\;y = \rho'\,y$ for all $y\not=x$.
\end{prop}

The first additional property above states the so-called ``substitution lemma,'' connecting the interpretation of a substituted term to the  interpretation of the original term in an updated environment---thus, roughly speaking, connecting syntactic and semantic substitution. The second additional property states that the interpretation of a term is oblivious to how its fresh (non-free) variables are evaluated.

\subsection{The two-sorted syntax of $\lambda$-calculus with values emphasized} \label{sub-inst-CBV}

We can split the syntax of $\lambda$-calculus in two syntactic categories, by distinguishing the subcategory of 
values, which consist of variables, constants and $\Lm$-terms.  This distinction is 
quite customary when modeling higher-order programming language semantics, where values are 
the only programs that have a ``static'' identity (whereas the non-values must be run/evaluated). 
Thus, we consider the mutually recursive syntactic categories of 
values, ranged over $V,W$ and (arbitrary) terms, ranged over by $X,Y,Z$: % (as before):    
$$
\begin{array}{rcl}
X    &\;::=\;& \InVl\;\, V   \;\mid\;   \App\;X\;Y
\\
V   &\;::=\;&  \Var\;x \;\mid\; \Ct\;c  \;\mid\; \Lm\;x\;X
\end{array}
$$
where $\InVl$ is the injection of values  
into terms.

We capture the above syntax by 
instantiating our signature to consist of 
two sorts and the desired constructors.  
Applying the same systematic deep-to-shallow transfer process as for the previous one-sorted syntax, 
we obtain:
\begin{myitem}
\item the ``native'' types $\valT$ and $\term$ for values and terms
\item the expected constructors, e.g., 
$\InVl : \valT \ra \term$
\item the standard operators, one for either syntactic 
category, e.g., $\fresh_{\valTT} : \var \ra \valT \ra \bool$ and 
$\fresh_{\termm} : \var \ra \term \ra \bool$.
\end{myitem} 
in what follows, we will omit the sort index 
for the operators, writing, e.g., 
$\fresh$ for both $\fresh_{\valTT}$ and 
$\fresh_{\termm}$. 

The framework-provided induction, recursion and semantic interpretation principles now refer to these mutually recursive types.  
Induction allows us to prove two simultaneous predicates and recursion/interpretation allows 
us two define two simultaneous functions, one on values and one on terms.
%---recursion for this syntax will be 
%illustrated in Section~\ref{sec-CBV}. 
%
For example, here are the corresponding instances of semantic domain and interpretation: 

\begin{defi} \label{def-sem}
	A {\em semantic domain} consists of two types, $\SSS$ and $\SSSv$, 
	endowed with the 
	functions $\invl : \SSSv \ra \SSS$, 
	$\app : \SSS \ra \SSS \ra \SSS$, 
	$\ct : \const \ra \SSSv$, 
	 and $\lm : (\SSSv \ra \SSS) \ra \SSSv$ 
	(corresponding to the term and value constructors $\InVl$, $\App$, $\Ct$ and $\Lm$). 
\end{defi}

\begin{prop}\label{prop-sem-val} 
		Let $(\SSS,\SSSv)$ %,\invl,\app,\ct,\lm)$ 
		be a semantic domain, and let $\val$ be the type of valuations of %the 
	variables in the semantic-value carrier of the domain, 
	$\var \ra \SSSv$. 
	Then 
	there exist the unique functions 
	$\semt : \term \ra \val \ra \SSS$ and 
	$\semv : \valT \ra \val \ra \SSS$ 	 
	such that:
	\begin{myitem}
		\item $\semt\;(\InVl\;V)\;\rho = \invl\;(\semv\;V\;\rho)$
		\item $\semt\;(\App\;X\;Y)\;\rho =
		\app\,(\semt\;X\;\rho)\,(\semt\;Y\;\rho)$ 
		\item $\semv\;(\Var\;x)\;\rho = \rho\;x$
		\item $\semv\;(\Ct\;c)\;\rho = \ct\;c$ 
		\item $\semv\;(\Lm\;x\;X)\;\rho = \lm\,(\lambda s.\;\semt\;X\;(\rho[x \la s]))$  
	\end{myitem} 
	\par
	In addition, the interpretation satisfies the following properties: 
	\begin{myitem}
		\item $\semt\;(X[V\,/\,y])\;\rho = \semt\;X\;(\rho[y \la \semv\;V\;\rho])$
		\item $\semv\;(W[V\,/\,y])\;\rho = \semv\;W\;(\rho[y \la \semv\;V\;\rho])$
		\item $\fresh\;x\;X$ and  $\rho=_{x}\rho'$ imply $\semt\;X\;\rho = \semt\;X\;\rho'$ 
		\item $\fresh\;x\;V$ and  $\rho=_{x}\rho'$ imply $\semv\;V\;\rho = \semv\;V\;\rho'$
	\end{myitem} 
\end{prop}
 
Note that this particular syntax has {\em two} sorts of terms ($\lambda$-calculus terms and values) and {\em one} sort of variables. Consequently, we have {\em two} semantic interpretation functions parameterized by {\em one} type of valuations.

\section{Call-By-Name $\lambda$-Calculus}
\label{sec-CBN}

In this section, we show how we have used our framework's infrastructure to formalize 
%two landmark 
some results in the theory of call-by-name (CBN) $\lambda$-calculus. 
We start with defining the CBN $\beta$-reduction relation (Section \ref{sub-CBN-Reds}) and 
proving its soundness with respect to 
the semantic interpretation of terms in  Henkin-style models (Section \ref{sub-CBN-sound}).
We continue with proving the Church-Rosser  
theorem \cite{bar-lam}, which states that the order in which CBN redexes are reduced is irrelevant ``in the long run'' (Section \ref{sub-CBN-CR}). 
Then, in a more substantial technical development, we prove 
the standardization theorem \cite{plotkin-CBNandCBVandLambda}, which states 
that reducibility is not restricted if we impose 
a canonical reduction strategy, based on identifying left-most redexes (Section \ref{sub-CBN-Std}).  
Finally, we develop and prove adequate a simple HOAS encoding---of 
$\lambda$-calculus into itself (Section \ref{sub-CBN-HOAS}). 
In each case, we emphasize the use of our framework's various features to leverage the formalization. 

All throughout this section, we employ the (single-sorted) syntax of $\lambda$-calculus 
with constants described in Section~\ref{sub-inst-CBN}.   
Following Plotkin \cite{plotkin-CBNandCBVandLambda}, we also fix a partial function $\constapp$ that shows how to apply a 
constant $c_1$ to another constant $c_2$; $\constapp\;c_1\;c_2$ can  
be either $\None$, meaning ``no result,'' 
or $\Some\;X$, meaning ``the result is $X$.''

\subsection{Call-by-name $\beta$-reduction} \label{sub-CBN-Reds}

Evaluation of a $\lambda$-calculus term proceeds by reducing {\em redexes}, which 
are subterms of one of the following two kinds:
\begin{myitem} 
	\item either {\em $\beta$-redexes}, of the form $\App\;(\Lm\;y\;X)\;Y$, which are reduced 
	to $X\,[Y \,/\, y]$
	\item or {\em $\delta$-redexes}, of the form $\App\;(\Ct\;c_1)\;(\Ct\;c_2)$ such that 
	$\constapp\;c_1\;c_2$ has the form $\Some\;X$,
	 which are reduced to $X$  	
\end{myitem}
The first are general-purpose redexes arising when an abstraction meets an application, 
whereas the second are  custom redexes representing the functionality built in 
the constants.

In the CBN calculus, there is no restriction on the terms $Y$ located at the right of 
$\beta$-redexes, reflecting the intuition that the argument $Y$ is passed to the function 
$\Lm\;y\;X$ ``by name,'' i.e.,  without first evaluating it. 
This style of reduction is captured by the following definition:

\begin{defi} \label{def-CBN-red}
The {\em one-step (CBN) reduction} relation $\!\!\redS\!\! : \term \ra \term \ra \bool$ 
is defined inductively by the following rules:\vspace{5pt}
\[
\begin{array}{ccc}
%&& \\
\dfrac{}{\App\ (\Lm\ y\ X)\ Y \redS X\ [Y\ /\ y]}\ \ \ \ \mbox{\rm ($\beta$)}
&\ \ \ \ \ \ \ \ 
&
\dfrac{\constapp\ c_1\ c_2\ =\ \Some\ X}{\App\ c_1\ c_2\redS X}
\ \ \ \ \mbox{\rm ($\delta$)}
\ \\[10pt]
%&& \\
\dfrac{X\redS X'}{\App\ X\ Y\redS \App\ X'\ Y}\ \ \ \ (\mbox{\rm \textsf{AppL}})
&\ \ \ \ \ \ \ \ &
\dfrac{Y\redS Y'}{\App\ X\ Y\redS \App\ X\ Y'}\ \ \ \ \mbox{\rm (\textsf{AppR})}\\[10pt]
%&& \\
\dfrac{X\redS X'}{\Lm\ y\ X \redS \Lm\ y\ X'}\ \ \ \ \mbox{\rm ($\xi$)}
&\ \ \ \ \ \ \ \ &
\end{array}
\]\\
The reflexive-transitive closure of $\!\redS\!$, denoted by $\!\MredS\!\!$, is called {\em multi-step reduction}. The equivalence closure $\!\redS\!$, denoted by $\beq\,$, is called {\em $\beta$-equivalence}. 
\end{defi}

Above, the rules \textsf{(AppL)}, \textsf{(AppR)} and $(\xi)$ delve into 
the term to locate a redex, whereas $(\beta)$ and $(\delta)$ perform its reduction. Note that $X \redS X'$ means that $X'$ was obtained from 
$X$ by the reduction of {\em precisely one} 
(nondeterministically chosen) redex.

\subsection{Soundness of $\beta$-equivalence with respect to  Henkin-style models}  \label{sub-CBN-sound}

As discussed in Section \ref{sub-inst-CBN}, our framework's notion of semantic domain is generic to any binding syntax. In particular cases, it yields 
meaningful semantic concepts after suitable customization.
For example, if we instantiate the framework to first-order logic and choose the semantic operators properly, we obtain the standard notion of first-order model with the Tarskian satisfaction relation \cite[\S6]{blanchette-frocos2013}. 

For our syntax of interest, a different kind of customization is necessary. In order to obtain Henkin-style standard notions of  set-theoretic models
for the $\lambda$-calculus \cite{bar-lam,mit-fou,hin-lam,DBLP:journals/iandc/Meyer82}, we do not need to choose particular semantic operators, but only to axiomatize their behavior. 
As an example, we pick one such notion, called {\em environment model} in \cite{DBLP:journals/iandc/Meyer82}.   

\begin{defi}  
 An {\em environment model} is a tuple 
 $(\SSS,\ct,\app,\lm,\ValidFuns)$ where 
  $(\SSS,\ct,\app,\lm)$ is a semantic domain and $\ValidFuns \su (\SSS \ra \SSS)$ is a set of functions such that  following hold: 
 \begin{description}
 	\item{\rm (1)} $\constapp\;c_1\;c_2 = \Some\;c$ 
 	implies 
 	$\app\,(\ct\;c_1)\,(\ct\;c_2) = \ct\;c$
 	\item{\rm (2)} $f \in \ValidFuns$ implies  
 	$\app\,(\lm\;f) = f$
 	\item{\rm (3)} 
 	$\lambda s.\;\sem\;X\;(\rho[x \la s]) \in \ValidFuns$
 \end{description} 
\end{defi} 

We think of the functions in $\ValidFuns$ as those that represent valid semantic behavior of functions induced by $\lambda$-terms. The three conditions express that (1) the semantic constants behave like the syntactic ones, (2) $\app$ is the left inverse of $\lm$ on valid functions (the semantic version of $\beta$) and (3) certain term-induced functions are valid. The motivation for condition (3) is the standard one in Henkin-style semantics: It ensures that the recursively defined semantic interpretation (Prop.~\ref{prop-sem}) employs valid  functions in the $\Lm$-case.  

With our available infrastructure, 
the formal statement and proof of the soundness theorem is easy: 

\begin{thm} 
	\label{th-CBN-sound}
Let $(\SSS,\ct,\app,\lm,\ValidFuns)$ be an environment model and let $\sem$ be its corresponding 
interpretation function. 
%(described in  Prop.~\ref{prop-sem}).
Then   
$X \beq Y$ implies $\sem\;X = \sem\;Y$.  
\end{thm} 

The theorem follows from the soundness of one-step reduction, i.e., the fact that $X \redS Y$ implies $\sem\;X = \sem\;Y$. 
The proof of the latter goes by rule induction on the definition of $\!\!\redS\!\!$ (Def.~\ref{def-CBN-red}). The substitution lemma (built in our framework as the last-but-one point of Prop.~\ref{prop-sem}) plays a key role when dealing with the 
($\beta$) case. Here is the standard argument, cast in our framework: We must prove 
$$\sem\,(\App\,(\Lm\;y\;X)\;Y)\,\rho =  \sem\,(X\;[Y\;/\;y])\,\rho
$$ 
To this end, we apply the Prop.~\ref{prop-sem} clauses for $\App$, $\Lm$ and substitution, which reduces our goal to 
$$\app\,(\lm\,(\lambda s.\,\sem\;X\;(\rho[y \la s])))\,(\sem\;Y\;\rho) = 
\sem\;X\;(\rho[y \la \sem\;Y\;\rho])$$ 
The last is true by points (2) and (3) of the environment model definition. 

In conclusion, our framework's %offers some %basic 
infrastructure facilitates the formalization of statements about the semantic interpretation of syntax.

\subsection{The Church-Rosser theorem} \label{sub-CBN-CR}

A binary relation $\succ$ is called {\em confluent} provided it satisfies the following ``diamond'' 
property: For all $u,v_1,v_2$ such that $u \succ v_1$ and $u \succ v_2$, 
there exists $w$ such that $v_1 \succ w$ and $v_2 \succ w$. In other words, every span can be joined. 
The Church-Rosser theorem states that this is the case for multi-step reduction: 

\begin{thm} %[Church-Rosser] 
	\label{th-CBN-CR}
	$\!\MredS\!\!$ is confluent. 
\end{thm}

%Thus, the theorem states that, if we select no %matter how we select 
%the redexes for reduction in two different ways, 
%there is always the possibility to converge by later reductions. 

A difficulty when trying to prove this theorem is the need to work with multiple reduction steps. 
Indeed, $\!\!\redS\!\!$ itself is not confluent, as seen by the following example, where we use 
the standard $\lambda$-calculus notation ($\lambda$ for abstraction, juxtapostion for application, etc.). Let $X =
(\lambda\;x_1.\,x_1\,x_1)\;X_1
 %\App\;(\Lm\;x_1\;(\App\;(\Var\;x_1)\;(\Var\;x_1)))\;X_1
 $, where 
$X_1 = (\lambda\,x.\,x)\;c
%\App\;(\Lm\;x\;(\Var\;x))\;(\Ct\;c)
$. 
If we choose to reduce the top redex of $X$, we obtain 
$X \redS Y_1$, where $Y_1 =
(x_1\;x_1)\,[X_1 \,/\, x_1] = X_1\;X_1
 %(\App\;(\Var\;x_1)\;(\Var\;x_1))\,[X_1 \,/\, x_1] = \App\;X_1\;X_1
$. 
On the other hand, if we choose to reduce the inner redex of $X$ (within $X_1$), 
we obtain 
$X \redS Y_2$, where $Y_2 = 
(\lambda\;x_1.\, x_1\;x_1)\;c
 %\App\;(\Lm\;x_1\;(\App\; (\Var\;x_1)\; (\Var\;x_1)))\;(\Ct\;c)
$. 
In order to join $Y_1$ and $Y_2$, intuitively we must perform the complementary reductions: 
By reducing the top redex in $Y_2$, we obtain $Y_2 \redS Z$, where $Z = c\;c %\App\;(\Ct\;c)\;(\Ct\;c)
$.  
However, $Y_1$ is not just one, but two redexes away from $Z$, meaning that 
$Y_1 \redS Z$ does not hold (although $Y_1 \MredS Z$ does). 

Dealing with multiple steps in the proof is possible, but the reasoning becomes %quite 
intricate. 
A more elegant solution, due to William Tait, 
 proceeds along the following lines \cite{bar-lam}:
 \vspace*{-1ex}
\begin{description}
	\item{(1)} First define a relation $\!\redP\!$ allowing the reduction of multiple (zero or more) redexes in parallel
	and prove that its transitive closure, $\!\MredP\!\!$, is the same as $\!\MredS\!\!$. 
	\item{(2)} Then prove that $\!\redP\!$ is confluent---which should be possible thanks to parallelism. 
	In the above example, we would have $Y_1 \redP Z$ by the parallel reduction of two $Z$-redexes. 
\end{description}
 \vspace*{-0.3ex}
Then the proof of the Church-Rosser theorem would be immediate: Since $\!\redP\!$ is confluent, 
than so is $\!\MredP\!\!$, i.e., $\!\MredS\!\!$. Next we proceed with tasks (1) and (2). 

\begin{defi}
	\label{def-redp-cbn}
The {\em one-step parallel reduction} relation $\!\!\redP\!\! : \term \ra \term \ra \bool$ 
is defined inductively by the following rules:\vspace{5pt}
\[
\begin{array}{ccc}
%&& \\
\dfrac{\constapp\ c_1\ c_2\ =\ \Some\ X}{\App\ c_1\ c_2 \redP X}
\ \ \ \ \mbox{\rm ($\delta$)}
&\ \ \ \ \ \ \ \ 
& 
\dfrac{X \redP X'\ \hspace*{3ex} Y \redP Y'}{\App\ (\Lm\ y\ X)\ Y \redP X'[Y'\ /\ y]}\ \ \ \ 
\mbox{\rm ($\beta$)}
\\[10pt]
%&& \\
\dfrac{X \redP X'\ \hspace*{3ex} Y \redP Y'}{\App\ X\ Y \redP \App\ X'\ Y'}\ \ \ \  \mbox{(\rm \textsf{App})}
&\ \ \ \ \ \ \ \ &
\dfrac{
	X \mbox{ \rm \small has the form } \Var\;x \mbox{ \rm\small or } \Ct\;c
}{X \redP X}\ \ \ \ \mbox{(\rm \textsf{Refl})}\\[10pt] 
%&& \\
\dfrac{X \redP X'}{\Lm\ y\ X \redP \Lm\ y\ X'}\ \ \ \ \mbox{\rm ($\xi$)}
&\ \ \ \ \ \ \ \ &
\end{array}
\]
%\\
%By taking the reflexive-transitive closure of it we obtain the \textbf{CBN parallel reduction}, ``$\_\MredP\_$''.
\end{defi}

The key technical differences between the definition of $\!\!\redP\!\!$ and that of 
$\!\!\redS\!\!$ are the following. 
$\!\!\redS\!\!$ has distinct left and right rules for application, \textsf{(AppL)} and \textsf{(AppR)}, which (together with $(\xi)$) navigate towards the single redex to be targeted for reduction via the $(\beta)$ rule, which is a base case. 
By contrast, 
$\!\!\redP\!\!$ deals with the immediate subterms $X$ and $Y$ of terms $\App\;X\;Y$ in parallel, through two alternative routes:
\begin{myitem}
	\item either by processing both subterms, via the \textsf{(App)} rule 
	\item or, if the term happens to form a redex,  optionally 
	reducing that top redex {\em and} 
	processing both subterms, via the $(\beta)$ rule (which is no longer a base case)
\end{myitem}
In addition, $\!\!\redP\!\!$ has a reflexivity rule, \textsf{(Refl)}, which deals with the idle components of the term (those not affected by reduction). \textsf{(Refl)} only applies to variables and constants, but it could have been allowed to apply to arbitrary terms, to the same effect:
\begin{lemma}\label{lem-refl-par}
	$X \redP X$ holds for any term $X$. 
\end{lemma} 

It is not difficult to prove (by standard rule induction, using Lemma~\ref{lem-refl-par}) that $X \redS Y$ implies $X \redP Y$  
and that $X \redP Y$ implies $X \MredS Y$, which ensure that $\!\MredP\! = \!\MredS\!$. 
This concludes task (1). Our formal proof required no special binding-aware type of reasoning, but  only  
%plain
standard inductive definitions and rule-induction proofs.   

Moving on to task (2), proving that $\!\redP\!$ is confluent, the simplest known approach 
is due to Takahashi   \cite{takahashi-CompleteDevelopment}. Let us assume  that $X \redP Y_1$ 
and $X \redP Y_2$, which means that both $Y_1$ and $Y_2$ have been obtained from $X$ 
by the parallel reduction of a number of redexes---it is the choice of 
which redexes have been reduced and which have been ignored (via the \textsf{(Refl)} rule)   
that constitutes the difference between $Y_1$ and $Y_2$. Hence, if $Z$ 
is the term obtained from $X$ by a {\em complete}  parallel reduction (with no redexes ignored)---which we write as $Z = \cdev\;X$---then $Z$ would be a valid join for $Y_1$ and $Y_2$. Indeed, $Z$ would be obtained 
from both $Y_1$ and $Y_2$ by reducing the redexes that had been ignored during the reductions of $X$ to $Y_1$ and $Y_2$.    

To define the complete parallel reduction operator (sometimes called ``complete development'' 
in the literature), $\cdev : \term \ra \term$, intuitively all we need to do is follow the 
inductive definition 
of parallel reduction and make that into a structurally 
recursive function---while restricting the application of the \textsf{(Refl)} rule 
to variables and constants only, for not skipping the reduction of any redex:
\[
\begin{array}{l}
\cdev\ (\Var\ x)=\Var\ x\ \ \ \ \ \ \ \ \ \cdev\ 
(\Ct\ c)=\Ct\ c\ \ \ \ \ \ \ \ \cdev\ (\Lm\ y\ X)=\Lm\ y\ (\cdev\ X)
\\\vspace*{-1.2ex}\\ 
\cdev\ (\App\ X\ Y)\ =\ \begin{cases}
\cdev\ Z,\ \ \text{if}\ (X,Y) \text{ have the form } (\Ct\;c_1,\,\Ct\;c_2) \\{\phantom{\cdev\ Z,\ \ \text{if}}}\text{ with } \constapp\;c_1\;c_2=\Some\;Z \\
(\cdev\ Z)\ [(\cdev\ Y)/ y],\ \ \ \ \text{if}\ X \text{ has the form }\Lm\ y\ Z\\
\App\ (\cdev\ X)\ (\cdev\ Y),\ \ \ \ \text{otherwise}
\end{cases} 
\end{array}
\]

However, the problem is that this definition is not {\em a priori} guaranteed 
to be correct, given that terms are not a free datatype due to quotienting to alpha-equivalence. 
One approach would be to redefine $\cdev$ on (unquotiented) quasi-terms and prove that it respects 
alpha-equivalence, but this would be technically quite difficult and would require 
breaking the term abstraction layer. Our recursion principle provides a better alternative: 
The above clauses are almost sufficient to construct an FSw model. What we additionally 
need is a specification of the expected behavior of the to-be-defined $\cdev$ with respect to 
freshness and swapping---which is straightforward, since $\cdev$ is expected to 
preserve freshness:  
$$\fresh\ y\ X\ \mbox{ implies }\fresh\ y\ (\cdev\ X)$$
and commute with swapping: 
 $$\cdev\ (X[z_1\ \swap\  z_2])=(\cdev\ X)[z_1\ \swap\ z_2].$$ 

Our recursion principle can now be employed to produce the following definition:

\begin{prop}\label{def-cdev}
	$\cdev : \term \ra \term$ is the unique function satisfying all the above clauses.  
(for the term constructors as well as the freshness and swapping operators). 
\end{prop}

Indeed, rewriting these clauses to make the required structure on the target type 
explicit, we see that they simply state 
the commutation of $\cdev$ with the constructors and the operators    
as described in Prop.~\ref{prop-CBN-rec}, where: 
\begin{myitem}
\item $\VAR=\Var$ and $\CT=\Ct$
\item $\LM\ x\ X'\ X\ =\ \Lm\ x\ X$
\item $\APP\ X'\ X\ Y'\ Y\ =\ 
\begin{cases}
Z\ \ \text{if}\ (X',Y') \text{ have the form } (\Ct\;c_1,\,\Ct\;c_2) \text{ with } \constapp\;c_1\;c_2=\Some\;Z \\
Z\ [Y / y]\ \ \ \ \text{if}\ X \text{ has the form }\Lm\ y\ Z\ \text{ and }\ X' \text{ has the form }\Lm\ y'\ Z'\\
\App\ X\ Y\ \ \ \ \text{otherwise}
\end{cases}
$ 
\item $\FRESH\ x\ X'\ X\ =\ \fresh\ x\ X$
\item $\SSWAP\ X'\ X\ z_1\ z_2\ =\ X[z_1\ \swap\ z_2]$
\end{myitem}
%Below we define these operators we obtain the function with the desired properties by recursion (theorem \ref{prop-CBN-rec}); we don't state explicitly the Horn clauses F1-F5, Sw1-Sw4 and FSw that render these, however these are proved 
Verifying the FSw model clauses for the above 
is completely routine. (Again, the desired facts follow by Isabelle's  ``auto'' proof method, which in this case applies 
the natural simplification rules for term constructors and operators.)   
%Note that Prop.~\ref{prop-CBN-rec} does not require the target type of the 
%defined function to be a ``syntactic'' domain such as $\term$ (or to satisfy any finite-support property)---although this happens to be 
%the case here.
%a quite common case, e.g., when defining syntactic translations. 
%
With the definition of $\cdev$ in place, it remains to prove the following: 

\begin{lemma} \label{lemma-cdev}
$X\redP X'$ implies $X'\redP \cdev\ X$
\end{lemma}

The informal proof of this lemma would go by induction on $X$, applying the 
Barendregt convention in the $\Lm$-case, i.e., when $X$ has the form $\Lm\;y\;Y$, 
to ensure that the bound variable $y$ is fresh for $X'$. One might expect that 
the structural fresh induction principle (Prop.\ \ref{th-CBN-fresh-ind}) is ideal for formalizing this 
task. However, the problem is that $\cdev$ analyzes $X$ more than one-level 
deep---when testing if $X$ is a $\beta$-redex, i.e., 
has the form $\App\;(\Lm\;x_1\;X_1)\;X_2$. This means that, in an inductive proof, we know that the fact holds for $X_1$ and $X_2$ and must prove that it holds for 
 $\App\;(\Lm\;x_1\;X_1)\;X_2$---this goes one notch beyond structural induction. We therefore use 
induction on the depth of $X$,  
and take advantage of Barendregt's variable convention 
by means of the fresh case distinction principle (Prop.\ \ref{th-CBN-fresh-case}) instead.

\subsection{The standardization theorem} \label{sub-CBN-Std}

The relation $\!\redS\!$ %and $\!\MredS\!\!$ 
makes a completely nondeterministic choice of the redex it reduces. 
The standardization theorem \cite{plotkin-CBNandCBVandLambda} refers to enforcing, without loss of expressiveness, 
a ``standard'' reduction strategy, which 
prioritizes leftmost redexes. %First, we define the left reduction relation: 

\begin{defi} \label{def-redL-cbn}
	The {\em one-step left reduction} relation $\!\!\redL\!\! : \term \ra \term \ra \bool$ 
	is defined inductively by the following rules:\vspace{5pt}
	\[
	\begin{array}{ccc}
	%&& \\
	\dfrac{\constapp\ c_1\ c_2\ =\ \Some\ X}{\App\ c_1\ c_2 \redL X}
	\ \  \mbox{(\rm \textsf{$\delta$})}
	&\ \ \ \  &
	\dfrac{}{\App\ (\Lm\ y\ X)\ Y \redL X\ [Y\ /\ y] }\ \ \ \ \mbox{(\rm \textsf{$\beta$})}\\[10pt]
	\dfrac{X \redL X'}{\App\ X\ Y \redL \App\ X'\ Y}\ \  \mbox{(\rm \textsf{AppL})}
	&\ \ \ \   &
	\dfrac{X \mbox{ \rm \small has the form } \Var\;x \mbox{ \rm\small or } \Ct\;c \;\;\;\;\;\;\;\;\; Y \redL Y'}{\App\ X\ Y \redL \App\ X\ Y' }\ \  \mbox{(\rm \textsf{AppR})} 
	\end{array}
	\]
	%\\
	%By taking the reflexive-transitive closure of it we obtain the \textbf{CBN reduction}, ``$\_\MredL\_$''.
\end{defi}

A first difference between $\!\!\redL\!\!$ and $\!\!\redS\!\!$ is that the former 
gives preference to redexes located towards the lefthand side of the term---as shown by the fact that the rule \textsf{(AppL)} has no restriction on $Y$, 
whereas \textsf{(AppR)} requires $X$ to be a variable or a constant. 
In other words, 
exploring the righthand side of the term in search for redexes is only allowed 
if exploring the lefthand side is no longer possible. 
Another difference is that $\!\!\redL\!\!$ does not reduce under $\Lm$---as shown by the absence of a 
\textsf{($\xi$)} rule.

\begin{defi} The {\em standard reduction (s.r.) sequence} predicate $\srs : \term\;\llist \ra \bool$ 
is defined inductively by the following rules:
\[
\begin{array}{ccccc}
%&& \\
\dfrac{}{\srs\;[\Ct\ c]}
\ \  \mbox{(\rm \textsf{Ct})}
&\ \   &
\dfrac{}{\srs\;[\Var\ x]}
\ \  \mbox{(\rm \textsf{Var})}
&\ \   &
\\[10pt]
\dfrac{X\ \redL\ \hd\;\Xs\;\;\;\;\;\;\;\srs\;\Xs}{\srs\;(X \cdot \Xs)}\ \  \mbox{(\rm \textsf{Red})}
&\ \    &
\dfrac{\srs\;\Xs}{\srs\;(\map\;(\Lm\;x)\;\Xs)}\ \  \mbox{(\rm \textsf{Lm})} 
&\ \   &
\dfrac{\srs\;\Xs\;\;\;\;\;\;\srs\;\Ys}
{\srs\;(\zipApp\;\Xs\;\Ys)}\ \  \mbox{(\rm \textsf{App})}
\end{array}
\]
\end{defi}

Above, for any $a$, $[a]$ denotes the singleton list containing $a$ and 
$\hd$, $\cdot$ and $\map$ denote the usual head, append and map functions on lists.
Moreover, $\zipApp$ applied to two lists $[X_1,\dots,X_n]$ and $[Y_1,\dots,Y_m]$ 
yields the list $[(\App\ X_1\ Y_1,\,\dots,\,\App\ X_n\ Y_1,\,\dots\,\,,\App\ X_n\ Y_m)]$ 
(obtained from first applying to $Y_1$ the terms $X_1,\ldots,X_n$, followed 
by applying $X_n$ to the terms $Y_2,\ldots,Y_m$).

A standard reduction sequence $[X_1,\dots,X_n]$ represents a systematic 
way of performing reduction, prioritizing left reduction, but also eventually exploring rightward located redexes. 
Thus, the rule \textsf{(App)} merges two 
s.r.\ sequences under the $\App$ constructor, 
scheduling the left one first and the right one second. 
The standardizaton theorem states that standard reduction sequences cover all 
possible reductions.

\begin{thm} %[Standardization] 
	\label{th-CBN-std}
	$X \MredS X'$ iff there exists a s.r.\ sequence 
	starting in $X$ and ending in $X'$. 
\end{thm}

The ``if'' direction, stating that 
s.r.\ sequences are subsumed by arbitrary reduction sequences, follows immediately by rule induction on the definition of $\srs$. 
So let us focus on the ``only if'' direction. It turns out that it is easier to 
use the multi-step parallel reduction $\!\MredP\!\!$ instead of $\!\MredS\!\!$---which is 
OK since we know from Section~\ref{sub-CBN-CR} that they are equal. To have better control 
over $\!\!\redP\!\!$ (and over $\!\!\MredP\!\!$), we need to be able to count the number of redexes 
that are being reduced in a step $X \redP Y$. In his informal proof, Plotkin
defines this number by a recursive traversal of the derivation tree for $X \redP Y$. 
Since we defined the relation $\!\!\redP\!\!$ inductively, i.e., as a least fixed point, 
we do not have direct access to the derivation trees. Instead, we introduce this number 
in a labeled variation of $\!\!\redP\!\!$, defined inductively as follows:

\begin{defi} \label{defi-labeledPar}
	The {\em labeled one-step parallel reduction} relation $\!\!\redP_{\!\!\!\_}\! : \term \ra \term \ra \nat \ra \bool$ 
	is defined inductively by the following rules:\vspace{5pt}
	\[
	\begin{array}{ccc}
	%&& \\
	\dfrac{\constapp\ c_1\ c_2\ =\ \Some\ X}{\App\ c_1\ c_2 \lredP{1} X}
	\ \  \mbox{\rm ($\delta$)}
	&\ \ \
	& 
	\dfrac{X \lredP{m} X'\ \hspace*{3ex} Y \lredP{n} Y'}{\App\ (\Lm\ y\ X)\ Y 
		\lredP{\scriptsize \colorbox{lightgray}{${\scriptsize 1 + m + n * \textsf{\rm no}\;X'\;y}$}} X'[Y'\ /\ y]}\ \ 
	\mbox{\rm ($\beta$)}
	\\[16pt] 
	%&& \\
	\dfrac{X \lredP{m} X'\ \hspace*{3ex} Y \lredP{n} Y'}{\App\ X\ Y \lredP{m+n} \App\ X'\ Y'}\ \  \mbox{(\rm \textsf{App})}
	&\ \ \   &
	\dfrac{
		X \mbox{ \rm \small has the form } \Var\;x \mbox{ \rm\small or } \Ct\;c
	}{X \lredP{0} X}\ \  \mbox{(\rm \textsf{Refl})}\\[13pt]
	%&& \\
	\dfrac{X \lredP{m} X'}{\Lm\ y\ X \lredP{m} \Lm\ y\ X'}\ \  \mbox{\rm ($\xi$)}
	&\ \ \ \  &
	\end{array}
	\]
	%\\
	%By taking the reflexive-transitive closure of it we obtain the \textbf{CBN parallel reduction}, ``$\_\MredP\_$''.
\end{defi}

The definitional rules for $\!\!\redP_{\!\!\!\_}\!$ are identical to those for $\!\!\redP\!\!$, 
except that they also track the number of reduced redexes. This number evolves as expected, 
e.g., for applications the left and right numbers are added. The most interesting 
rule is that for $\beta$-reduction, where the label of the conclusion is 
$1 + m + n * \textsf{\rm no}\;X'\;y$. 
This is obtained by counting:
\begin{myitem}  
\item $1$ for the top redex (which is being explicitly reduced in the rule)
\item $m$ for the redexes being reduced in $X$ to obtain $X'$ 
\item $n * \textsf{\rm no}\;X'\;y$ 
for the $n$ redexes being reduced in $Y$ to obtain $Y'$,  
{\em one set for each (free) occurrence of $y$ in $X'$}---because the occurrences of $y$ in $X'$ correspond to the  
occurrences of $Y$ in $X'[Y/y]$ 
that will be reduced to $Y'$ 
\end{myitem}
(We recall that $\textsf{\rm no}\;X'\;y$ counts the number of (free) occurrences of the variable $y$ in $X'$, via the operator $\textsf{\rm no}$ defined at the end of Section~\ref{sub-inst-CBN}.) 
% 
%Back to the proof of the theorem,
 
Now, using an easy lemma stating that $X \redP Y$ is equivalent to  
the existence of $n:\nat$ such that $X \lredP{n} Y$, we are left with proving the following:

\begin{prop} %[Standardization] 
	\label{th-CBN-std-aux}
	If $X \MlredP{m} X'$, then there exists a s.r.\ sequence 
	starting in $X$ and ending in $X'$. 
\end{prop}

The proof idea for the above is to build the desired s.r.\ sequence 
by ``consuming'' $X \MlredP{n} X'$ one step at a time, from left to right, as expressed below: 

\begin{prop}\label{th-CBN-par-srn}
If $X \lredP{m} X'$ and $\Xs$ is a s.r.\ sequence starting in $X'$, then 
there exists a s.r.\ sequence starting in $X$ and ending in the last term of $\Xs$. 
\end{prop}

Prop.~\ref{th-CBN-par-srn} easily implies Prop.~\ref{th-CBN-std-aux} by rule induction 
on the definition of the reflexive-transitive closure; in the base case, one uses the fact that 
$\src\;[X]$ holds for all terms $X$, which follows immediately by rule induction on the definition of of $\src$. 

So it remains to prove Prop.~\ref{th-CBN-par-srn}. 
The proof requires a quite elaborate induction, namely lexicographic induction on three measures: 
the length of $\Xs$, the number (of $X$-to-$X'$ reduction steps) $m$ and the depth of $X$.  
Inside the induction proof, there is a case distinction on the form of $X$. 

The most complex case is when $X$ is an application, since here we have to deal with 
the redexes. For handling the $\beta$-redex subcase, two lemmas are required. 
The first states that $\!\!\redP_{\!\!\!\_}\!\!\ $ preserves substitution, while keeping 
the numeric label under a suitable bound:  

\begin{lemma} \label{subst-par}
	If $X \lredP{m} X'$ and $Y \lredP{n} Y'$, then there exists 
	$k$ such that $k \leq m + \no\;X'\;y * n$ and $X\,[Y\,/\,y]  \lredP{k} X'\,[Y'\,/\,y]$. 
\end{lemma}

It is proved by induction on the depth of $X$, making essential use of the 
property that connects $\no$ with substitution, which is built in our definition of $\no$ 
(Def.~\ref{no-def}).   
The second expresses commutation between (labeled) parallel reduction and left reduction:   

\begin{lemma} \label{pl-commute}
	If $X \lredP{m} Y$ and $Y \redL Z$, then 
there exist $Y'$ and $n$ such that $X \MredL Y'$ and $Y' \lredP{n} Z$. 
\end{lemma}

It is proved by lexicographic induction on $m$ and the depth of $X$. 
% (employing a few auxiliary lemmas).
%
Back to the proof of Prop.~\ref{th-CBN-par-srn}, the other cases (different from $\App$) are 
conceptually quite straightforward. However, the formal treatment of the $\Lm$-case raises a 
subtle issue, which we describe next. 

The informal reasoning in the $\Lm$-case goes as follows: Assume $X$ has the form $\Lm\;y\;Y$. 
Then, for inferring $\Lm\;y\;Y \lredP{m} X'$, the last applied rule must have been 
either \textsf{(Refl)} or \textsf{($\xi$)}. In the case of \textsf{(Refl)}, we have $X=X'$ 
so the desired s.r.\ sequence is $\Xs$. In the case of \textsf{($\xi$)}, 
\coll{\mbox{we obtain that $X' = \Lm\;y\;Y'$ for some $Y'$ such that $Y \lredP{m} Y'$}}. 
Moreover, since $\Xs$ is a s.r.\ sequence starting in $\Lm\;y\;Y'$, there must be 
a s.r.\ sequence $\Ys$ starting in $Y'$ such that $\Xs = \map\;(\Lm\;y)\;\Ys$. 
By the induction hypothesis, we obtain a s.r.\ sequence $\Ys'$ starting in $Y$ and 
ending in the last term of $\Ys$. Hence we can take $\map\;(\Lm\;y)\;\Ys'$ to be 
the desired s.r.\ sequence (starting in $X$).

The above informal argument applies (among other things) a special inversion rule 
for $\!\!\lredP{\_}\!\!$, taking advantage of knowledge about the shape of the 
lefthand side of the conclusion: a term of the form $\Lm\;y\;Y$. However, 
as emphasized above, it is implicitly assumed that an application of the \textsf{($\xi$)} rule 
with $\Lm\;y\;Y$ as lefthand side of its conclusion will have the 
form 
$$\dfrac{Y \lredP{m} Y'}{\Lm\;y\;Y \lredP{m} \Lm\;y\;Y'}$$ 
i.e., will ``synchronize'' 
with the variable $y$ bound in $Y$. In other words, we need the following inversion rule: 

\begin{lemma} \label{lem-inv-spec}
	If $\Lm\;y\;Y \lredP{m} X'$, then one of the following holds:
	\begin{myitem}
		\item $X' = \Lm\;y\;Y$ (meaning \textsf{(Refl)} must have been applied)
		\item There exists $Y'$ such that $X' = \Lm\;y\;Y'$ and $Y \lredP{m} Y'$ 
		(meaning a $y$-synchronized \textsf{($\xi$)} must have been applied)
	\end{myitem}
\end{lemma}

Proving the above is not straightforward, and relies on some properties of $\!\lredP{m}\!\!$ that are global, i.e., depend on the behavior of its rules different from \textsf{($\xi$)}.   
All we can get from the standard 
inversion rule (coming from the inductive definition of $\!\lredP{m}\!\!$) is, in the second case, the existence of $z$, $Z$ and $Z'$  
such that $\Lm\;y\;Y = \Lm\;z\;Z$, $X' = \Lm\;z\;Z'$ and $Z \lredP{m} Z'$. 
Using the properties of equality between $\Lm$-terms, we obtain that $Y = Z\,[y \swap z]$. 
To complete the proof of Lemma \ref{lem-inv-spec}, we further need the following: 

\begin{lemma}\label{lem-equivar}
$\!\!\lredP{\_}\!\!$ \ is equivariant, i.e., 
$Z \lredP{m} Z'$ implies $Z\,[y \swap z] \lredP{m} Z'\,[y \swap z]$. 	
\end{lemma}

\begin{lemma}\label{lem-ffresh}
	$\!\!\lredP{\_}\!\!$ \ preserves freshness, i.e., 
$\fresh\;y\;Z$ and $Z \lredP{m} Z'$ implies $\fresh\;y\;Z'$. 	
\end{lemma}

Using these lemmas and the basic properties of freshness and swapping, we define 
$Y'$ to be $Z'\,[y \swap z]$ and obtain $\Lm\;y\;Y' = \Lm\;z\;Z'$ and $Y \lredP{m} Y'$; 
in particular, $X' = \Lm\;y\;Y'$ and $Y \lredP{m} Y'$, as desired. This concludes our outline 
of the proof of Prop.~\ref{th-CBN-par-srn} and overall of the standardization theorem.   

%\vspace*{0.5ex}
%\noindent
%{\bf A note on formal infrastructure. } We have proved corresponding 
%versions of Lemmas~\ref{lem-equivar}, \ref{lem-ffresh} and \ref{lem-inv-spec} for all 
%our one-step reduction relations---which was useful since the aforementioned reasoning 
%pattern occurred multiple times in proofs for these different relations.  
%Moreover, for all of them we proved a fresh rule induction principle in the style of 
%Nominal logic \cite{urban-Barendregt} (essentially the rule-induction counterpart 
%of Prop.~\ref{th-CBN-fresh-ind}). 

\subsection{Adequate HOAS encoding}  \label{sub-CBN-HOAS}

Next we describe another case study, which takes advantage of our framework's increased substitution-awareness: the formal definition and proof of an adequate HOAS encoding of CBN $\lambda$-calculus into itself. The technique we describe 
here would also apply to more complex encodings in logical frameworks. 

\vspace*{1ex}
\noindent
{\bf HOAS encoding of syntax.} 
A %(so far irrelevant) 
feature of our formalized syntax of 
$\lambda$-calculus is that the type $\const$ of constants is not fixed; rather,  
the type $\term$ is parameterized by an unspecified type $\const$. This is captured in Isabelle as a polymorphic type. 
%in  a type variable $\const$.  
The feature has not been very important so far, but becomes crucial for our HOAS application. We will use two instances of this polymorphic type: 
\begin{myitem}
	\item 
one as before, with constants from a type $\const$, which we still denote by $\term$, and 
\item one with constants from $\const \cup \{\ctapp,\ctlm\}$ (i.e., $\const$ enriched with 
two new constants, $\ctapp$ and $\ctlm$, corresponding to the term constructors $\App$ and $\Lm$), which we denote by $\cterm$
\end{myitem}

Switching to standard $\lambda$-notation for a moment, the natural HOAS encoding of
$\term$ in $\cterm$ should be a %recursive function $\enc : \term \ra \cterm$ 
characterized by the following equations:
\begin{description}
	\item{\rm (1)} $\enc\;x = x$
	\item{\rm (2)} $\enc\;c = c$
	\item{\rm (3)} $\enc\,(X\;Y) = \ctapp\;(\enc\;X)\,(\enc\;Y)$
	\item{\rm (4)} $\enc\,(\lambda x.\,X) = \ctlm\;(\lambda x.\;\enc\;X)$
\end{description}
In our formalization, these equations are:
\begin{description}
	\item{\rm (1)} $\enc\,(\Var\;x) = \Var\;x$
	\item{\rm (2)} $\enc\;(\Ct\;c) = \Ct\;c$
	\item{\rm (3)} $\enc\,(\App\;X\;Y) = \App\,(\App\;\ctapp\,(\enc\;X))\,(\enc\;Y)$
	\item{\rm (4)} $\enc\,(\Lm\;x\;X) = \App\;\ctlm\;(\Lm\;x\;(\enc\;X))$
\end{description}

Two central properties of HOAS encodings are preservation of freshness and commutation with substitution, the latter usually called {\em compositionality} \cite{har-fra,Pfenning01computationand}---here is their statement for our case:
\begin{description}
	\item{\rm (5)} $\fresh\;x\;X$ implies 
	$\fresh\;x\;(\enc\;X)$
	\item{\rm (6)} $\enc\,(X[Y\,/\,y]) = 
	(\enc\;X)[(\enc\;Y)\,/\,y]$  
\end{description}

As usual, the problem with the equations (1)--(4) is that they are not guaranteed to be valid on alpha-equated terms. Our framework again offers an immediate resolution via Prop.~\ref{prop-CBN-rec}: In exchange for some trivial term properties to check, 
it provides a function $\enc$ satisfying not only (1)--(4), but also (5) and (6). 

\begin{defi}\label{defi-hoas-enc}
	$\enc : \term \ra \cterm$ is the unique function satisfying clauses (1)--(6). 
\end{defi}

In fact, here we have an example where Prop.~\ref{prop-rec-extra} applies too, offering us two additional facts about $\enc$ (again, in return for the verification of some trivial properties of terms):
\begin{description}
	\item{\rm (7)} $\enc$ is injective 
	\item{\rm (8)} The ``iff'' version of clause (5) holds 
\end{description}

Clauses (6) and (7) form what is usually called the {\em (syntactic) adequacy} property of a HOAS encoding.\footnote{In typed frameworks, the adequacy property additionally ensures that the encoding is a bijective correspondence between the terms of the original system and some canonical forms in the host system.} 
One could also argue that (8), which is 
seldom stated explicitly in the HOAS literature, 
 should be verified as well in order to deem an encoding adequate.  
Our framework's recursion principle seems almost specialized in delivering such adequacy ``packages.'' 

Here are the aforementioned basic properties that we have been required to check in order for Prop.~\ref{prop-CBN-rec} and \ref{prop-rec-extra} to apply, guaranteeing the above properties of $\enc$. 
The clauses (1)--(6) indicate the following FSb model structure having carrier type $\term'$.  
The constructor-like functions are $\Var$, $\Ct$, the function mapping $X$, $X'$, $Y$, $Y'$ to $\App\,(\App\;\ctapp\;X)\;Y$, and the function mapping $x$, $X$, $X'$ to 
$\App\;\ctlm\;(\Lm\;x\;(\enc\;X))$. 
Note that these last two functions ignore 
the ``primed'' arguments (members of $\term$); this is because only iteration is needed here (rather than full-fledged recursion). 
The freshness- and substitution-like operators are the usual $\fresh$ and $\_[\_/\_]$, again ignoring the primed arguments. 

The fact that the above forms an FSb model amounts to the following: 
\begin{description}
	\item{F1:} $\fresh\ x\ (\Ct\ c)$
	\item{F2:} $x\not=z$ implies $\fresh\;z \;(\Var\;x)$ 
	\item{F3:} $\fresh\;z\;X$ and $\fresh\;z\;Y$ implies $\fresh\;z\;(\App\;(\App\;\ctapp\;X)\;Y)$ 
	\item{F4:} $\fresh\;x\;(\App\;\ctlm\;(\Lm\;x\;X))$
	\item{F5:} $\fresh\;z\;X$ implies $\fresh\;z\;(\App\;\ctlm\;(\Lm\;x\;(\enc\;X)))$
	\item{Sb1:} $(\Var\;z)[Z\,/\,z] = Z$
	\item{Sb2:} $x\not=z$ implies 
	$(\Var\;x)[Z\,/\,z] = \Var\;x$
	\item{Sb3:} $(\App\;(\App\;\ctapp\,X)\;Y)\,[Z'\,/\,z] = 
	\App\;(\App\;\ctapp\,(X[Z'\,/\,z]))\;
	(Y[Z'\,/\,z])$ 
	\item{Sb4:} $x \not=z$ and $\fresh\;x\;Z$ implies 
	$(\App\;\ctlm\;(\Lm\;x\;X))\,[Z\,/\,z]  = \App\;\ctlm\;(\Lm\;x\;(X[Z\,/\,z]))$
	%%%%
	\item{SbRn:} $x\not=y$ and $\fresh\;y\;X$ implies 
	$\App\;\ctlm\;(\Lm\;y\;(X[(\Var\;y)\, / x])) = 
	\App\;\ctlm\;(\Lm\;x\;X)$ 
\end{description} 
The fact that the model is freshness-reversing amounts to the following:
\begin{description} 
	\item{F2c:} $\fresh\;z \;(\Var\;x)$ implies $x\not=z$  
	\item{F3c:} $\fresh\;z\;(\App\,(\App\;\ctapp\;X)\;Y)$ implies 
	$\fresh\;z\;X$ and $\fresh\;z\;Y$  
	\item{F4$\_$5c:} $\fresh\;z\;(\App\;\ctlm\;(\Lm\;x\;X))$ 
	implies $x=z$ or $\fresh\;z\;X$
\end{description}
The fact that the model is constructor-injective amounts to the aforementioned constructor-like functions being injective and non-overlapping. 

All the above follow immediately 
(and are proved in Isabelle automatically)
from the  standard properties of substitution and freshness---commutation with the term constructors, our framework stores as proved lemmas. For example, facts F1--F5 and their converses follow from the standard simplification facts for freshness w.r.t.\ the term constructors, and SbRn follows from 
Prop.~\ref{prop-CBN-quasi-inj}(2) and the injectivity of $\App$.  
%(Our framework has simplification rules that cover the clauses F1--F5 in the ``iff'' version, which cover 
%F2c, F3c and F4$\_$5c as well.) 

\vspace*{1ex}
\noindent
{\bf HOAS encoding of the reduction relation.} 
So far, we have used the $\term'$ syntax to adequately encode the $\term$ syntax. 
In order to be able to encode inductively defined relations on $\term$, we will need to organize $\term'$ as miniature logical framework. 
Unlike in full-fledged logical frameworks such as Edinburgh LF \cite{har-fra} or Generic Isabelle \cite{pau-genTh}, it will not have its own built-in mechanism for specifying logics or calculi---instead, we will use the ``external'' mechanism of inductive definitions of relations over $\term'$. 
%
%To achieve an encoding in the spirit of LF and Generic Isabelle, we will need to consider a background notion of term equivalence. For example, LF employs explicit proofs inhabiting types that represent judgments, and the latter are identified modulo $\beta$ or $\beta\eta$. Similarly, Generic Isabelle uses $\beta\eta$-equivalence. 
%
The background term equivalence will be  $\beta$-equivalence, $\beq\,$. 

With these provisions, we can encode inductively defined $n$-ary relations $R$ on $\term$ as inductively defined $n$-ary relations $R_{\textsf{h}}$ on $\term'$, where:
\begin{myitem}
\item Each inductive clause in the definition of $R$ is matched by an inductive clause in the definition of $R_{\textsf{h}}$. 
\item There is an additional ``background'' clause in the definition of $R_{\textsf{h}}$ that states compatibility with $\beta$-equivalence.
\end{myitem}

All the relations on $\term$ defined in this paper can be encoded in this manner. As an example we choose the left reduction relation 
$\!\!\redL\!\!$, which will be encoded as a relation $\!\!\redLL\!\!$. 

\begin{defi} \label{def-redL-cbnH}
	The relation $\!\!\redLL\!\! : \term' \ra \term' \ra \bool$ 
	is defined inductively by the following rules:\vspace{0pt}
	\[
	\begin{array}{c}
	%&& \\
	\dfrac{}{\App\ (\App\ \ctapp\ (\App\;\ctlm\ X))\ Y \redLL X\;Y}\ \ \ \ \mbox{\rm ($\beta$')}
	\hspace*{7ex}
	\dfrac{\constapp\ c_1\ c_2\ =\ \Some\ X}{\App\ c_1\ c_2\redLL X}
	\ \ \ \ \mbox{\rm ($\delta$')}
	\\[13pt]
	%&& \\
	\dfrac{X\redLL X'}{
		\App\ (\App\ \ \ctapp\ X)\ Y \redLL 
		\App\ (\App\ \ \ctapp\ X')\ Y	
	}\ \ \ \ (\mbox{\rm \textsf{AppL'}})
	\\[13pt]
	\dfrac{X \mbox{ \rm \small has the form } \Var\;x \mbox{ \rm\small or } \Ct\;c \;\;\;\;\;\;\;\;Y\redLL Y'}{
		\App\ (\App\ \ \ctapp\ X)\ Y \redLL 
		\App\ (\App\ \ \ctapp\ X)\ Y'
	}
	\ \ \ \ \mbox{\rm (\textsf{AppR'})}
	\\[13pt]
	\dfrac{  
		X \beq Y
		\hspace*{4ex} Y \redLL Y' 
		    \hspace*{4ex}
		Y' \beq X'
	}
    {
		X \redLL X' 
    }
	\ \ \ \ \mbox{\rm (\textsf{Compat}$_\beq$)}
	\end{array}
	\] 
\end{defi}

The difference between the above clauses for $\!\!\redLL\!\!$ and the corresponding ones that define $\!\!\!\redL\!\!$ (in Def.~\ref{def-redL-cbn}) is that now $\Lm$ and $\App$ are employed as part of the meta-level infrastructure, whereas the object-level behavior of the application and abstraction constructors is tagged with the constants $\ctapp$ and $\ctlm$. The object-calculus substitution 
in rule $(\beta)$ is replaced by mere meta-level application in rule $(\beta')$. 
The background rule ({\textsf{Compat}$_\beq$)}
is responsible for ``fixing'' this mismatch 
between $(\beta)$ and $(\beta')$: The meta-level application of encoded items will be part of a $\beta$-redex, 
which is $\beta$-equivalent to a meta-level term obtained by applying meta-level substitution. This means that, ultimately, the object-level substitution in $(\beta)$ will correspond to meta-level substitution. 

Let us illustrate the above phenomenon, switching for a moment to standard $\lambda$-calculus notation. In this notation, the $(\beta)$ rule for $\!\!\redL\!\!$ is $(\lambda y.\;X)\,Y \redL X[Y/x]$, and the $(\beta')$ rule for 
$\!\!\redLL\!\!$ is $\ctlm\;X\;Y \redLL X\;Y$. An instance of $(\beta)$ is 
$(\lambda x.\;x)\,y \redL x[y/x]$, i.e., 
$(\lambda x.\;x)\,y \redL y$. The corresponding instance of $(\beta')$ is 
$\ctapp\;(\ctlm\;(\lambda x.\;x))\;y \redLL (\lambda x.\;x)\;y$. The two instances are related as follows:
\begin{myitem}
	\item $\enc\,((\lambda x.\;x)\,y) = \ctapp\;(\ctlm\;(\lambda x.\;x))\;y$, i.e., 
	the encoding of the lefthand side of the first
	is 
	the lefthand side of the second 
	\item $\enc\;y = y \beq (\lambda x.\;x)\;y$, i.e., 
	the encoding of the lefthand side of the first
	 is $\beta$-equivalent to 
	 the righthand side of the second	 
\end{myitem}

This suggests a statement of the adequacy of the encoding of $\!\!\redL\!\!$ as $\!\!\redLL\!\!$.  

\begin{thm} \label{thm-operAde}
	The following hold:
	\begin{description}
		\item{\rm (1)} If $\,X \redL Y$ then $\enc\;X \redLL \enc\;Y$. 
		\item{\rm (2)} If $\,\enc\;X \beq X'$ and $X' \redLL Y'$, then there exists $Y$ such that $X \redL Y$ and $\enc\;Y \beq Y'$.
		\item{\rm (3)} $X \redL Y$ iff $\enc\;X \redLL \enc\;Y$. 
	\end{description}
\end{thm}

Point (1) follows by rule induction on the definition of $\!\!\redL\!\!$. All cases are completely routine, except for that 
of the $(\beta)$ rule. In that case (using again standard $\lambda$-calculus notation for readability),  
we must prove $\enc\,((\lambda y.\;X)\;Y) \redL \enc\,(X[Y/y])$. We have the following, using $(\beta')$ and the properties of $\enc$, including compositionality: 
$$
\begin{array}{c}
\enc\,((\lambda y.\;X)\;Y) = 
\ctapp\;
(\ctlm\;(\lambda y.\;\enc\;X))\;
(\enc\;Y) \redLL 
(\lambda y.\;\enc\;X)\;(\enc\;Y)
\beq 
\\
\beq 
(\enc\;X)[(\enc\;Y) \,/\, y] 
=
\enc\,(X[Y/y])
\end{array}
$$
From this, using {(\textsf{Compat}$_\beq$)} 
we obtain 
$
\enc\,((\lambda y.\;X)\;Y)  
\redLL 
\enc\,(X[Y/y])
$, 
as desired.  
%
%Thus, to handle the HOAS encoding of $(\beta)$ we must use not only$(\beta')$, but also {(\textsf{Compat}$_\beq$)} and compositionality of the syntactic encoding.   

Point (2) follows by rule induction on the definition of $\!\!\redLL\!\!$, using some inversion rules of $\beq$ w.r.t. the syntactic constructors.   
 Point (3) has one implication covered by point (1). For the other implication,  
 we use point (2) and 
the following simple but crucial observation:  
\begin{lemma}
	$\enc\;X$ is a $\beta$-normal form (in that, for all $Y$, $\enc\;X \MredS Y$ implies $Y = \enc\;X$). 
\end{lemma}  
This ensures that $\enc\;X \beq \enc\;Y$ implies 
$\enc\;X = \enc\;Y$, which further implies $X=Y$ (by the injectivity of $\enc$). In turn, this immediately allows to prove (3)'s reverse implication from point (2).
%This lemma is used, together with the Church-Rosser theorem, for cutting down the nondeterminism introduced by the {(\textsf{Compat}$_\beq$)} rule. 

This concludes our formal exercise of deploying our framework for adequately encoding both syntax 
and reduction of CBN $\lambda$-calculus in a miniature HOAS framework. In the future, it will be interesting to explore the formalization of more complex frameworks using the same techniques.

\section{Call-By-Value $\lambda$-Calculus}
\label{sec-CBV}

The call-by-value (CBV) $\lambda$-calculus differs from the CBN 
$\lambda$-calculus by the insistence that only 
values are being substituted for variables in terms, i.e., 
a term is evaluated to a value before being 
substituted. All the notions pertaining to the CBV calculus are defined as a variation of their CBN counterparts 
by factoring in the above value restriction. 
The $\constapp$ partial function is now assumed to return values instead of arbitrary terms. 

\begin{defi} \label{def-CBV-red}
	The {\em one-step CBV reduction} relation $\!\!\vredS\!\! : \term \ra \term \ra \bool$ 
	is defined inductively by rules similar to those of 
	Def.~\ref{def-CBN-red}, namely by the rules (\textsf{AppL}) and (\textsf{AppR}) from 
	there (of course, with $\!\!\vredS\!\!$ replacing $\!\!\redS\!\!$), together with:  
	%\pagebreak  %%% typesetting hack
%	
  \[
  \begin{array}{l}
  \dfrac{}{\App\ \coll{(\InVl\ (\Lm\ y\ X))}\ \coll{(\InVl\  W)}\ \vredS\ X\ [ W\ /\ y]}\ \ \ \ (\beta)
  \end{array}
  \] 
  \vspace*{-2ex}
	\[
	\begin{array}{ccc}
	%&& \\
	\dfrac{\constapp\ c_1\ c_2\ =\ \Some\ \coll{V}}{\App\ c_1\ c_2\ \vredS\ \coll{\InVl\  V}}
	\ \ \ \ (\delta)
	&\ \ \ \ \ \ \ \ &
	%\dfrac{X\ \vredS\ X'}{\App\ X\ Y\ \vredS\ \App\ X'\ Y}\ \ \ \ (\textit{appL})
	%\\[8pt]
	%&& \\
	%\dfrac{Y\ \vredS\ Y'}{\App\ X\ Y\ \vredS\ \App\ X\ Y'}\ \ \ \ (\textit{appR})
	%&\ \ \ \ \ \ \ \ &
	\dfrac{X\ \vredS\ X'}{\coll{\InVl\ (\Lm\ y\ X)}\ \vredS\ \coll{\InVl\ (\Lm\ y\ X')}}\ \ \ \ (\xi)
	\\[10pt]
	\end{array}
	\] 
\end{defi}

Highlighted above are the differences between the one-step CBV reduction and its CBN counterpart.  
In the (\textit{$\delta$}) and (\textit{$\xi$}) rules the differences are inessential: One 
employs the value-to-term injection $\InVl$ to account for the fact that $\constapp$ 
returns a value and that $\Lm$-terms are values. 
%Notice that we have two types involved 
%now and therefore it is necessary to employ the injection $\InVl$ when we want to take some value into full terms.
The essential difference shows up in the 
(\textit{$\beta$}) rule, which requires the righthand side of the redex to be a value. 
Similar differences are highlighted in the next definitions.

\begin{defi} \label{def-redp-cbv}
The {\em one-step parallel CBV reduction} relation $\!\!\vredP\!\! : \term \ra \term \ra \bool$ 
is defined inductively by rules similar to those of 
Def.~\ref{def-redp-cbn}, namely by the rules (\textsf{App}) and (\textsf{Refl}) from 
there (with $\!\!\vredP\!\!$ replacing $\!\!\redP\!\!$), together with:\vspace{5pt}
\[
\dfrac{\constapp\ c_1\ c_2\ =\ \Some\ \coll{V}}{\App\ c_1\ c_2 \vredP \coll{\InVl\ V}}
\ \ \mbox{\rm ($\delta$)}
\hspace*{6ex}
\dfrac{X \vredP X'\ \hspace*{3ex} Y \vredP \coll{\InVl\ V'}}{\App\ \coll{(\InVl\ (\Lm\ y\ X))}\ Y \vredP X'[V'\ /\ y]}\ \ \
\mbox{\rm ($\beta$)} 
\]
\vspace*{-2ex}
\[
\dfrac{X \vredP X'}{\coll{\InVl\ (\Lm\ y\ X)} \vredP \coll{\InVl\ (\Lm\ y\ X')}}\ \ \ \ \mbox{\rm ($\xi$)}
\]
%\\
%By taking the reflexive-transitive closure of it we obtain the \textbf{CBN parallel reduction}, ``$\_\MredP\_$''.
\end{defi}

%\vspace*{1ex}

\begin{defi} \label{def-redL-cbv}
	The {\em one-step left CBV reduction} relation $\!\!\vredL\!\! : \term \ra \term \ra \term$ 
	is defined inductively by rules similar to those of 
	Def.~\ref{def-redL-cbn}, namely by the rule (\textsf{AppL}) from 
	there (with $\!\!\vredL\!\!$ replacing $\!\!\redL\!\!$), together with:\vspace{5pt}
	\[ 
	\dfrac{\constapp\ c_1\ c_2\ =\ \Some\ \coll{V}}{\App\ c_1\ c_2 \vredL \coll{\InVl\ V}}
	\ \  \mbox{(\rm \textsf{$\delta$})}
	\hspace*{6ex}
	\dfrac{}{\App\ \coll{(\InVl\ (\Lm\ y\ X))}\ \coll{(\InVl\  W)} \vredL X\ [W\ /\ y] }\  \mbox{(\rm \textsf{$\beta$})}\\[10pt]
	\]
	\vspace*{-3.5ex}
 \[
	\dfrac{Y \vredL Y'}{\App\ \coll{(\InVl\ V)}\ Y \vredL \App\ \coll{(\InVl\ V)}\ Y' } \mbox{(\rm \textsf{AppR})}  
	\]
	%\\
	%By taking the reflexive-transitive closure of it we obtain the \textbf{CBN reduction}, ``$\_\MredL\_$''.
\end{defi}

Except for the above definitions, the CBV concepts are identical to those of the CBN concepts, {\em mutatis mutandis}, i.e., plugging in the above CBV basic relations instead of the CBN ones. These include the multi-step versions of the relations and the notions of complete parallel reduction operator and standard reduction sequence. 

Moreover, the statements and proofs of the Church-Rosser and standardization theorems are essentially identical, {\em mutatis mutandis}. 
Like Plotkin has suggested in his informal development \cite{plotkin-CBNandCBVandLambda},  
the formal proofs could be easy adapted from CBN to CBV, obtaining: 

\begin{thm} \label{th-CBV-CR-Std}
Theorem \ref{th-CBN-CR} and Theorem \ref{th-CBN-std} hold with the same statements, 
after replacing the CBN notions with their CBV counterparts.  
\end{thm}

While the CBN and CBV formal developments are conceptually very similar, for the latter we employed our framework's infrastructure for a two-sorted syntax. 
To illustrate how 
this two-sorted syntax is handled by the framework, we show the definition of the CBV counterpart of $\cdev$. (We omit the sort annotation, $\term$ or $\valT$, form the substitution and swapping operators.) %Def.~\ref{def-CBN-cdev}.  

\begin{defi}\label{def-CBV-cdev} 
The \emph{CBV complete parallel reduction operator} of a term $X$ (written $\ \cdev_{\termm}\  X$) and of a value $ V$ (written $\ \cdev_{\valTT}\  V$) are the unique pair of functions satisfying:
\[
\begin{array}{l}
\cdev_{\valTT}\ (\Var\ x)=\Var\ x\ \ \ \ \ \ \ \ \ \cdev_{\valTT}\ 
(\Ct\ c)=\Ct\ c
\\ \vspace*{-1.2ex}\\
%\ \ \ \ \ \ \ \ 
\cdev_{\termm} \ (\InVl\  V) =\InVl\ (\cdev_{\valTT}\ V)\ \ \ \ \ \ \ \ \cdev_{\valTT}\ (\Lm\ y\ X)=\Lm\ y\ (\cdev_{\termm} \ X)
\\ \vspace*{-1ex}\\
\cdev_{\termm} \ (\App\ X\ Y)\ =\ \begin{cases}  
\InVl\,(\cdev_{\valTT} \ V),\\
\hspace*{3ex}
\text{if}\ (X,Y) \text{ have the form } (\InVl\,(\Ct\;c_1),\,\InVl\,(\Ct\;c_2)) \\
\hspace*{3ex}
\text{ with } \constapp\;c_1\;c_2=\Some\;V
%\InVl(\dres\ (\cdev_{\termm} \ X)\ (\cdev_{\termm} \ Y)),\\
%\ \ \ \ \ \ \ \ \text{if}\ (\App\ X\ Y)\ \text{is a $\delta$-redex}\\
\\
(\cdev_{\termm} \ Z)\ [(\cdev_{\valTT}\,W)/ y],\\
\hspace*{3ex}
\text{if}\ (X,Y) \text{ have the form } (\InVl\,(\Lm\ y\ Z),\,\InVl\  W)
\\
\App\ (\cdev_{\termm} \ X)\ (\cdev_{\termm} \ Y),\ \ \ \ \text{otherwise}
\end{cases}
\\ \vspace*{-1ex}\\
\fresh_{\valTT}\ y\  V\ \mbox{ implies }\ \fresh_{\valTT}\ y\ (\cdev_{\valTT}\  V)\ \ \ \ \ \ \\ \vspace*{-1.2ex}\\
\fresh_{\termm} \ y\ X\ \mbox{ implies }\ \fresh_{\termm} \ y\ (\cdev_{\termm} \ X)
\\ \vspace*{-1.2ex}\\
\cdev_{\valTT}\ ( V[z_1\ \swap\ z_2])=(\cdev_{\valTT}\  V)[z_1\ \swap\ z_2]
\\ \vspace*{-1.2ex}\\
\cdev_{\termm} \ (X[z_1\ \swap\ z_2] )=(\cdev_{\termm} \ X)[z_1\ \swap\ z_2]
\end{array}
\]
\end{defi}

Similarly to the CBN case, this turns out to be a correct definition thanks to a two-sorted version of Prop.~\ref{prop-CBN-rec}, 
that is, via exhibiting a two-sorted FSw model.

\section{Overview of the Formalization 
}
\label{sec-overview}

%As already mentioned, our development is based on a general theory of syntax with bindings,  
%which was presented elsewhere \cite{ghepop-2017-jar}. 
%The development has two parts. 

The formalization presented in this paper has two parts. 
The first part is the instantiation of the general theory to the two syntaxes, 
	of $\lambda$-calculus and of $\lambda$-calculus with emphasized values, together 
	with the transfer from a deep to a more shallow embedding---which produces all the ``infrastructure'' concepts and theorems reported in Section~\ref{sec-inst}. 
	This is currently a completely routine, but very tedious process: It spans over 
	more than 15000 lines of code (LOC) for each syntax. 
	The reasons for this large size are the sheer number of stated theorems about 
	constructors and substitution (more than 
	300 facts for the one-sorted syntax and more than 500 for the two-sorted syntax) 
	and the many intermediate facts stated in the process of transferring the recursion theorems. 
	Thanks to 
	using a custom template for the instantiation, the whole process only took us two person-days. 
	However, this is unreasonably long for a process that can be entirely automated---so we leave its automation as a pressing goal for future work.

The second part is the theory of CBN and CBV $\lambda$-calculus, culminating with the proofs 
of the soundness, Church-Rosser, standardization and HOAS adequacy theorems (reported in Sections \ref{sec-CBN} and \ref{sec-CBV}). 
This is where our routine effort from the first part fully paid off. Thanks to our comprehensive collection 
of facts about substitution and freshness, we were able to focus almost entirely on 
formalizing the high-level ideas present in the informal proofs---notably in Plotkin's sketches 
of his elaborate proof development for the standardization theorem.  
%On two occasions---for defining Takahashi's complete parallel reduction operator and the number of variable occurrences needed in %Plotkin's 
%the Standardization proof development---our recursion principle allowed us to quickly get off the ground, in the second case also offering useful freshness and substitution lemmas needed later in the proof.   
Altogether, the second part consists of 5500 LOC (2500 for CBN and 3000 for CBV) 
and took us one person-month. The appendix gives concrete pointers to the Isabelle formalization, including a map of the theorems listed in this paper and their formal counterparts. 

An exception to the above general phenomenon (of being able to focus on the high-level proof ideas)  
was the need to engage in the low-level task of proving custom constructor-directed 
inversion rules for our reduction relations---illustrated and motivated in the discussion leading to Lemma~\ref{lem-inv-spec}.  
This lemma is just 
one example of the 
several similar inversion rules we proved, corresponding to the  
inductive rules involving $\lambda$-abstraction in the reduction relations' definitions. 
%In fact, keeping in mind future developments, we have proved such rules even when they were not needed for our particular results.  
These rules are essentially the binding-aware version of what Isabelle/HOL offers 
via the ``inductive cases'' command \cite{isa-refman}. 
They seem to be generally useful in proof developments that involve inductively defined reductions but require induction over terms. 
Binding-aware inversion principles form an integral part of higher-order abstract syntax frameworks \cite{abellaJournalPaper,beluga,DBLP:journals/entcs/PoswolskyS09,DBLP:conf/cade/PfenningS99}, and have also been discussed (though unfortunately not implemented) in the context of Isabelle Nominal 
\cite{BerghoferU08_NominalInversionPrinciples}. 
%The literature on formal reasoning seems to have overlooked the general usefulness of these rules; and state of the art definitional packages such as Isabelle Nominal do not attempt to infer them automatically. 
%notwithstanding their amenability for automatic derivations. 

Finally, our 
case study   
%formal proof of the Standardization theorem 
illustrates another interesting and apparently 
not uncommon %largely overlooked 
phenomenon: that fresh structural induction on terms may be too weak in proofs, whereas depth-based induction in conjunction with fresh cases may do the job {\em while still enabling the use of  Barendregt's convention}---as illustrated in our proof of Lemma~\ref{lemma-cdev}.

\section{Related Work}
\label{sec-relWork}

This paper's contribution is twofold: 
(1) it instantiates our general framework to two particular syntaxes, showing how to deploy the framework's induction and recursion principles and (2) it performs two specific formal reasoning case studies for these syntaxes. We split the discussion of related work in two corresponding subsections. 

\subsection{Formal approaches to syntax with bindings}

There is a large amount of literature on formal approaches to syntax with bindings, many of which are supported by proof assistants or logical frameworks. 
(See \cite[\S2]{POPLmark}, \cite[\S6]{momFelty-Hybrid4} and
\cite[\S8]{ghepop-2017-jar} for overviews.) 
%
%Our line of work follows a {\em nameful} paradigm to representing binders %(\ref{subsec-rel-paradigm}) and takes a {\em universe}-like approach to capture syntax specified by an arbitrary binding signature (\ref{subsec-rel-univ}), which is restricted to single-variable binders but caters for infinitely branching terms (\ref{subsec-rel-arbSig}). We formalize a theory covering not only the syntactic constructors, but also other standard generic operators (\ref{subsec-rel-polytypic}), and featuring nominal-logic-style induction and operator-sensitive recursion (\ref{subsec-rel-reas}). 
%
%\subsection{Binding representation paradigm}\label{subsec-rel-paradigm}
%
These approaches roughly fall under three main paradigms
of reasoning about bindings.     
In the {\em nameful paradigm}, binding variables are
passed as arguments to the binding operator 
and terms are usually equated modulo alpha-equivalence. 
The best known 
rigorous account of this paradigm is offered by Gabbay and Pitts's 
nominal logic.   %\cite{DBLP:conf/lics/GabbayP99}. 
Originally developed within a non-standard axiomatization of set theory \cite{DBLP:conf/lics/GabbayP99,gabbayPittsNominal}, nominal logic was subsequently cast in a standard foundation 
\cite{pitts01nominal,pitts-AlphaStructural}, and also significantly developed in a proof assistant context---most extensively by Urban and collaborators %for Isabelle/HOL 
\cite{urban-NominalHOL,UrbanTasson,UrbanBerghof-RecCombNominal,urban-Barendregt,urbanGeneralBinders}.  
%  
%; its proximal formal competitor is the Nominal package \cite{}, which is also implemented in Isabelle. 
%
%

In the {\em nameless paradigm} originating with %the work of
De Bruijn \cite{bru-lam}, the bindings are indicated 
through nameless pointers to positions in a term.  
Major exponents of the scope-safe nameless paradigm are representations based on presheaves \cite{fio-abs,hof-sem} and nested datatypes  \cite{bird-DBnested,alt-reus}.   
The presheaf approach has been generalized and refined in many subsequent works, e.g., \cite{DBLP:conf/types/GambinoH03,Fiore08-cartesianClosedBi2008,indexedContainers,DBLP:conf/cpp/KaiserSS18,allais-bindingsByDependentTypes-agda,allais-icfp2018,DBLP:journals/iandc/HirschowitzM10}.     
%, which originated with the work of Pitts and Gabbay 
% 
%

Finally, the {\em higher-order abstract syntax (HOAS)} paradigm, 
\cite{phe-hig,har-fra,pau-genTh,DBLP:conf/cade/PfenningS99,weakHOAS,momFelty-Hybrid4,chlipala-Parametric,feltyPientka-comparison} based on ideas going back as far as 
Church \cite{Church-HOL}, 
Huet and Lang \cite{DBLP:journals/acta/HuetL78} and  
Martin-L\"{o}f  
\cite[Chapter 3]{Nordstrom:1990:PMT:92094}, 
has gained traction with the works of Harper et.~al \cite{har-fra}, Pfenning and Elliott \cite{pfenningOriginalHOAS} and Paulson~\cite{pau-genTh} in the late eighties. HOAS essentially embeds the binders of the represented system (referred to as the {\em object} system)  
shallowly into the meta-logic's binder.  
HOAS has been 
pursued in  
dedicated logical frameworks 
such as  
Abella \cite{abellaJournalPaper}, 
Beluga \cite{beluga}, 
Delphin \cite{DBLP:journals/entcs/PoswolskyS09} and
Twelf \cite{DBLP:conf/cade/PfenningS99}, and in general-purpose proof assistants such as 
Coq \cite{weakHOAS,chlipala-Parametric} and Isabelle \cite{gun-proper}.  
HOAS often allows for 
lighter formalizations, thanks to borrowing binding mechanisms and sometimes structural properties from the meta-level.
% Unary substitution is also 
%built in the representation:  
%term-for-variable substitution in strong HOAS and variable-for-variables substitution 
%in weak HOAS. 
% 
Formalizations in this paradigm are often accompanied by pen-and-paper proofs of the representations' adequacy (which 
involve informal reasoning about substitution) \cite{har-fra,Pfenning01computationand}; as shown in Section~\ref{sub-CBN-HOAS}, 
our substitution-aware recursion principle can ease
the formalization of such proofs.
Some approaches in the literature combine two paradigms. For example, the locally nameless approach  \cite{DBLP:conf/types/Pollack93,aydemirPOPL08,locallyNamelessOverview} employs a nameless representation of bindings, but stores a distinct type of variables that can occur free; 
this enables some essentially nameful techniques for dealing with free variables (similar to those of nominal logic). Other examples are the Hybrid system \cite{momFelty-Hybrid4} and the ``HOAS on top of FOAS'' approach \cite{pop-HOASOnFOAS}, which develop HOAS reasoning techniques over locally nameless and nameful representation substrata. 

Our work in this paper belongs to the nameful paradigm, giving a formal expression to many ideas from nominal logic---but departing from nominal logic through 
its focus on a rich built-in theory of substitution (including substitution-aware recursion) and built-in semantic interpretation.  
While our structural induction principle (Prop.~\ref{th-CBN-fresh-ind}) is essentially the same as the nominal logic one (as implemented in Coq \cite{nominalCoq} and Isabelle \cite{UrbanTasson}), our recursion principles (Prop.~\ref{prop-CBN-rec}) differ from the nominal logic one in two essential ways. First, our FSw-model-based principle, while factoring in freshness and swapping as primitives on the target domain like the nominal one, does {\em not} assume that the former is defined from the latter---this brings additional generality and has similarities to a principle formalized by Michael Norrish in HOL4 for the syntax of $\lambda$-calculus \cite{primrecFOAS-Norrish04}. Second, our FSb-model-based principle factors in substitution rather than swapping, which is arguably a more fundamental operator to syntax with bindings (notwithstanding the nominal logic's convincing case for the fundamental role of swapping). 
A current limitation of our recursion principles is their inability to handle freshness \emph{for parameters}. In particular, this means that we could not have used, say, our FSw-model-based principle to define substitution on (quotiented) terms. Instead, our framework performs a low-level definition of substitution on  (unquotiented) quasi-terms and then lifts it to terms. All these details are of course hidden from the user.

Our work seems to be the first to formalize generic support for 
the interpretation of terms in semantic domains---which in the meantime has also been developed in Agda 
within the well-scoped nameless paradigm, using a universe \cite{allais-icfp2018}. In the context of nominal logic, defining semantic interpretations  incurs some difficulties due to the absence of finite support \cite[page 492]{pitts-AlphaStructural}. 

Another difference between our approach and that of a definitional package such as Nominal Isabelle is that we statically verify the arbitrary-syntax meta-theory whereas they dynamically generate any instance of interest.  
For a more through %detailed
discussion of the distinguishing features of our general framework, 
including universe versus code-generator approaches, we refer the reader to \cite{ghepop-2017-jar}.     

In recent work \cite{BindingsAsFunctors}, we have made progress with integrating the definitional principles for syntax with bindings displayed in this paper with Isabelle/HOL's general-purpose definitional 
package for inductive and coinductive datatypes \cite{traytel-et-al-2012,blanchette-et-al-2014-tru,nonuniform-lics2017,DBLP:conf/esop/Blanchette0T15}  
enriching the recursion and corecursion \cite{fouco,amico} infrastructure with 
a binding-aware component. The setting of \cite{BindingsAsFunctors} is more general than that of this paper and of \cite{ghepop-2017-jar,our-own-paper}, since it allows for nesting and mixing types in flexible ways, and also leverages Isabelle/HOL's theory of cardinals \cite{cardHOL} to go not only beyond finite branching, but also beyond finite depth for terms with bindings (as with, e.g., B\"{o}hm trees \cite{bar-lam}).

\subsection{Similar case studies in other frameworks}

In a development that has become part of the Isabelle standard library, 
Nipkow and Berghofer \cite{isa-lambda,DBLP:conf/cade/Nipkow96} have proved 
several CBN $\lambda$-calculus properties, including Church-Rosser and Normalization. They use a de Bruijn encoding of $\lambda$-terms, which somewhat 
impairs the readability of their statements and proofs. 
The Isabelle Nominal package hosted many developments concerning (variants of) $\lambda$-calculus \cite{nomGr}, including the CBN Church-Rosser and standardization  \cite{isa-lambdaNom,DBLP:journals/corr/NageleOS16}, the second fixed point theorem \cite{ckhb-cpp11} and 
the meta-theory of Edinburgh's LF  \cite{urban-LFInNominal}.  

The Church-Rosser and standardization theorems have also been formalized in other provers: 
the Church-Rosser theorem in Abella \cite{cr-accattoli}, Coq \cite{cr-huet}, 
HOL \cite{homeierCR}, LEGO \cite{locallyNamed0}, 
PVS \cite{cr-shankar} and Twelf \cite{cr-pfenning}   
and the standardization theorem 
in Coq \cite{std-coquand} and LEGO  \cite{DBLP:conf/types/JuttingMP93,std-mckinna-pollack}. 
All of the above developments consider the call-by-name variant of $\lambda$-calculus (or of a more complex calculus)---which means our work provides the first formalization of these results for the call-by-value calculus. However, 
the call-by-value calculus has been formalized in other contexts, e.g., recently as a model of computation in Coq \cite{forster-smolka-lam-comp}.

Aspects of our framework's approach to semantic interpretation and HOAS encodings have already been presented in the second author's PhD thesis \cite[\S2.3]{pop-thesis} and in a previous conference paper \cite{pop-recPrin} (with some of the ideas going back to the work on term-generic logic \cite{DBLP:journals/tcs/0001R15}), but so far have not been developed as thoroughly as we do here. In particular, in this journal paper we cover environment models and the 
soundness of $\beta$-reduction and take a principled approach to adequacy of encodings in $\lambda$-calculus with constants and background $\beta$-reduction.  
The only other formalization of HOAS adequacy we are aware of is that of Cheney et al.~\cite{DBLP:journals/jar/CheneyNV12} using Nominal Isabelle, which covers a more complex case than ours: that of encoding $\lambda$-calculus in HOL. Admittedly, Nominal Isabelle already delivers well for the task of defining HOAS encodings and proving their adequacy. Yet, our framework seems able to target HOAS phenomena even more hands-on: It offers  the syntactic adequacy properties (including substitution compositionality and freshness preservation and reflection) as part of the recursion infrastructure, which leads to a very compact formulation and proof of adequacy. 

Apart from the novelty of some of the formalized results (e.g., concerning call-by-value), % concerning CBV, 
a main motivation for performing 
%our own formalization of these fundamental $\lambda$-calculus results 
these case studies is that they offered us the possibility to test essentially all our framework's features, from built-in substitution to induction and recursion principles to semantic interpretation to many-sortedness. 
We believe that these features have enabled us to produce a fully formal yet %high-level and 
pedagogical presentation of the results.  
%some fundamental $\lambda$-calculus results, which highlighted some new ideas concerning the relationship between informal and formal reasoning about bindings. 
In the future, it would be interesting to provide a comparison between our development and alternative developments in other frameworks.
%, as was done, e.g., in recent work with a case study on formalizing logical relation arguments \cite{POPLMarkReloaded-2019}. 

\subsection{Future work}

We plan to deploy our framework to formalize various aspects of HOL and Isabelle/HOL's metatheory  \cite{consIsa-2015,kp-esop-2017,us-t2s,DBLP:journals/pacmpl/Kuncar018,DBLP:conf/ictac/GengelbachW20}, complementing the work already done in the HOL4 prover on these aspects \cite{DBLP:conf/lpar/PohjolaG20}.

\vspace*{2ex}
\noindent
{\bf Acknowledgments.}  
Popescu has received funding from UK's Engineering and Physical Sciences Research Council 
(EPSRC) via the grant EP/N019547/1, Verification of Web-based  Systems (VOWS) 
and from VeTSS/NCSC through the grant ``Formal Verification of Information Flow Security for Relational Databases''.

\bibliographystyle{splncs03}
\bibliography{bib}{}

 \include{Appendix}

\end{document}

%% file: myCommands.tex
\newcommand\swap{\!\leftrightharpoons\!}

\newtheorem{thm}[lemma]{Theorem}
\newtheorem{prop}[lemma]{Prop}

\newtheorem{defi}[lemma]{Def}
\newcommand\leftOut[1]{}

\renewcommand{\phi}{\varphi}

\newcommand{\su}{\subseteq}

\renewcommand{\partial}{\rightharpoondown}
\newcommand{\la}{\leftarrow}

\newcommand{\ra}{\rightarrow}
\newcommand{\ValidFuns}{\mbox{\rm\textsf{\normalsize ValidFuns}}}

\newcommand{\sem}{\mbox{\rm\textsf{\normalsize sem}}}
\newcommand{\semv}{\mbox{\rm\textsf{\normalsize sem}}_{\valTT}}
\newcommand{\semt}{\mbox{\rm\textsf{\normalsize sem}}_{\termm}}

\newcommand{\enc}{\mbox{\rm\textsf{\normalsize enc}}}
\newcommand{\ctapp}{\mbox{\rm\textsf{\normalsize ctapp}}}
\newcommand{\ctlm}{\mbox{\rm\textsf{\normalsize ctlm}}}

\newcommand{\depthh}{\mbox{\rm\textsf{\normalsize depth}}}

\newcommand{\varsOf}{\mbox{\rm \textsf{\normalsize vars\hspace{-0.1ex}Of}}}
\newcommand{\None}{\mbox{\rm\textsf{\normalsize None}}}
\newcommand{\Some}{\mbox{\rm\textsf{\normalsize Some}}}

\newcommand{\Ct}{\mbox{\rm\textsf{\normalsize Ct}}}

\newcommand{\zipApp}{\mbox{\rm\textsf{\normalsize zipApp}}}

\newcommand{\InVl}{\mbox{\rm\textsf{\normalsize Val}}}

\newcommand{\srs}{\mbox{\rm\textsf{\normalsize srs}}}

\newcommand{\cdev}{\mbox{\rm\textsf{\normalsize cdev}}}
\newcommand{\no}{\mbox{\rm\textsf{\normalsize no}}}

\newcommand{\app}{\mbox{{\rm\textsf{\normalsize app}}}}
\newcommand{\ct}{\mbox{{\rm\textsf{\normalsize ct}}}}
\newcommand{\invl}{\mbox{{\rm\textsf{\normalsize val}}}}

\newcommand{\lm}{\mbox{{\rm\textsf{\normalsize lm}}}}

\newcommand{\Var}{\mbox{\rm\textsf{\normalsize Var}}}

\newcommand{\App}{\mbox{\rm\textsf{\normalsize App}}}

\newcommand{\Lam}{\mbox{\rm\textsf{\normalsize Lam}}}

\newcommand{\Lm}{\mbox{\rm\textsf{\normalsize Lm}}}

\newcommand{\Abs}{\mbox{\rm\textsf{\normalsize Abs}}}

\newcommand{\fresh}{\mbox{\rm\textsf{\normalsize fresh}}}
\newcommand{\FRESH}{\mbox{\rm\textsf{\normalsize FRESH}}}

\newcommand{\src}{\mbox{\rm\textsf{\normalsize src}}}

\newcommand{\map}{\mbox{\rm\textsf{\normalsize map}}}
\newcommand{\VAR}{\mbox{{\rm \textsf{\normalsize V$\hspace*{-0.1ex}$A$\hspace*{-0.1ex}$R}}}}
\newcommand{\APP}{\mbox{{\rm \textsf{\normalsize A$\hspace*{-0.1ex}$P$\hspace*{-0.1ex}$P}}}}
\newcommand{\LM}{\mbox{{\rm \textsf{\normalsize L$\hspace*{-0.1ex}$M}}}}
\newcommand{\CT}{\mbox{{\rm \textsf{\normalsize C$\hspace*{-0.1ex}$T}}}}
\newcommand{\SSWAP}{\mbox{\rm\textsf{\normalsize SWAP}}}
\newcommand{\SSUBST}{\mbox{\rm\textsf{\normalsize SUBST}}}

\newcommand{\Xs}{\mbox{\textit{Xs}}}

\newcommand{\Ys}{\mbox{\textit{Ys}}}

\newcommand{\cterm}{\bf term'}
\newcommand{\val}{\bf val}

\newcommand{\D}{\bf D}

\newcommand{\nat}{\mbox{\bf nat}}

\newcommand{\valT}{\mbox{\bf value}}
\newcommand{\valTT}{\mbox{\bf \scriptsize value}}

\newcommand{\bool}{\mbox{\bf bool}}
\newcommand{\const}{\mbox{\bf const}}

\newcommand{\var}{\mbox{\bf var}}

\newcommand{\term}{\mbox{\bf term}}
\newcommand{\termm}{\mbox{\bf \scriptsize term}}
\newcommand{\DDD}{\mbox{\bf D}}
\newcommand{\SSS}{\mbox{\bf S}}
\newcommand{\SSSv}{\mbox{\bf Sv}}

\newcommand{\abs}{\mbox{\bf abs}}

\newcommand{\param}{\mbox{\bf param}}
\newcommand{\option}{\mbox{\bf option}}

\newcommand{\llist}{\mbox{\bf list}}
\newcommand\thmcontinues[1]{Continued}

\newcommand{\sett}{\mbox{\bf set}}

\newcommand{\constapp}{\mbox{\rm\textsf{\normalsize Ctapp}}}

\newcommand{\hd}{\mbox{\rm\textsf{\normalsize hd}}}

\newcommand{\lredP}[1]{\mbox{$\ \Rightarrow_{#1}\ $}}
\newcommand{\MlredP}[1]{\mbox{$\ \Rightarrow_{#1}^*\ $}}

\newcommand{\beq}{\equiv}

\newcommand{\redP}{\mbox{$\ \Rightarrow\ $}} %one-step parallel reduction
\newcommand{\redS}{\mbox{$\ \rightarrow\ $}} %one-step standard reduction
\newcommand{\redLL}{\mbox{$\ \looparrowright_{\textsf{h}}\ $}}
\newcommand{\redL}{\mbox{$\ \looparrowright\ $}} %one-step left reduction
\newcommand{\MredP}{\mbox{$\ \Rightarrow^*\ $}} %parallel reduction
\newcommand{\MredS}{\mbox{$\ \rightarrow^*\ $}} %standard reduction
\newcommand{\MredL}{\mbox{$\ \looparrowright^*\ $}} %left reduction

\newcommand{\vredP}{\mbox{$\ \Rightarrow_{\textsf{v}}\ $}} %one-step parallel reduction
\newcommand{\vredS}{\mbox{$\ \rightarrow_{\textsf{v}}\ $}} %one-step standard reduction
\newcommand{\vredL}{\mbox{$\ \looparrowright_{\textsf{v}}\ $}} %one-step left reduction

%% file: Appendix.tex
\appendix

\noindent
{\Large APPENDIX}

\normalsize

\ \\ \ \\
The Isabelle theories can be downloaded from the paper's website \cite{lambda-scripts} and processed with Isabelle 2019. 
The general framework (applicable to an arbitrary syntax with bindings and reported in our companion paper \cite{ghepop-2017-jar}) is an entry in the Archive of Formal Proofs \cite{Binding_Syntax_Theory-AFP} and must be imported from there. Our development is based on that entry and is structured in three sessions (provided with their customary ROOT files \cite[\S2]{IsabelleSystemManual}): \textsf{Interface}, \textsf{Instance$\_$Lambda$\_$Syntax} 
and \textsf{Case$\_$Studies}. 

\subsection*{The \textsf{Interface} session}

This session pre-instantiates the general framework to several commonly encountered arities. The development is also syntax-independent, and can be regarded as being part of the general framework. 

\subsection*{The \textsf{Instance$\_$Lambda$\_$Syntax} session}

This session fully instantiate the framework to the two particular syntaxes discussed in this paper: the single-sorted (unsorted) one of $\lambda$-calculus (used for the CBN calculus) and the two-sorted variation that distinguishes values from other terms (used for the CBV calculus). It corresponds to Section~\ref{sec-inst}. 
The relevant theories in this session are called \textsf{L}, \textsf{L$\_$Inter}, \textsf{LV} and \textsf{LV$\_$Inter}.  

The theory \textsf{L} contains a wealth of facts that are made available for the (unsorted) syntax of $\lambda$-calculus after instantiating our framework (discussed in Section~\ref{sub-inst-CBN}). The theory file contains detailed comments to guide the reader through these facts. They cover properties of the constructors and the operators (freshness, swapping, unary substitution and parallel substitution), as well as induction and recursion and semantic-interpretation principles. 
The theory \textsf{LV} has a similar structure and content (though fewer comments), but considers the two-sorted syntax of $\lambda$-calculus with emphasized values (discussed in Section~\ref{sub-inst-CBV}). 

The theories \textsf{L$\_$Inter} and \textsf{LV$\_$Inter} further customize the two syntax instances with a few abbreviations and re-formulations of facts that we have deemed more convenient for this particularly simple syntaxes. Notably, they introduce the $\Lm$ constructor, which in \textsf{L$\_$Inter} has type $\var \ra \term \ra \term$, by putting together an abstraction constructor $\Abs : \var \ra \term \ra \abs$ and a one-binding-argument constructor, 
$\Lam : \abs \ra \term$. More precisely, $\Lm\;x\;X$ abbreviates $\Lam\;(\Abs\;x\;X)$. 
(Our general framework employs explicit abstractions as a separate syntactic category, whereas here we preferred to inline abstractions as part of a single $\Lm$-constructor.)

Here is a map between Section~\ref{sub-inst-CBN}'s propositions and their formal counterparts in theory \textsf{L}:\footnote{Note that the paper covers only a small subset of the facts provided in the formalization. The latter are best explored by reading the content of theory \textsf{L}, which includes detailed comments and explanations. The name of the operators and theorems follow a uniform pattern which can be understood by reading these comments.} 
\begin{itemize}
	\item Prop.~\ref{prop-CBN-quasi-inj} corresponds to lemmas \textsf{``Lam inj''} and 
	\textsf{``Abs$\_$lm$\_$lm swap$\_$vlm$\_$lm ex''} 
	\item Prop.~\ref{prop-CBN-subst_comp} 
	corresponds to lemmas \textsf{"subst$\_$vlm$\_$lm compose 1"} and \textsf{"subst$\_$vlm$\_$lm subst$\_$vlm$\_$lm compose 2"}
	\item Prop.~\ref{th-CBN-fresh-ind} corresponds to lemma \textsf{``induct fresh''} (reformulated as lemma \textsf{``induct fresh 2''} in theory \textsf{L$\_$Inter})
	\item Prop.~\ref{th-CBN-fresh-case}  
	corresponds to lemma \textsf{``term$\_$lm fresh cases''} (reformulated as lemma \textsf{``term fresh cases''} in theory \textsf{L$\_$Inter})
	\item Prop.~\ref{prop-CBN-rec} corresponds to lemmas \textsf{``wlsFSb rec term$\_$FSb$\_$morph''} and \textsf{ ``wlsFSw rec term$\_$FSw$\_$morph''}
	\item Prop.~\ref{prop-rec-extra} 
	corresponds to lemmas  
	\textsf{``wlsFSb rec refl$\_$freshAll''} and \textsf{``wlsFSb rec is$\_$injAll''}
	\item Prop.~\ref{prop-sem} corresponds to lemma 
	\textsf{``wlsSEM semInt comp$\_$int''}
\end{itemize}

\subsection*{The \textsf{Case$\_$Studies} session}

This session
	 contains the four case studies 
	 described in Sections~\ref{sub-CBN-sound}--\ref{sub-CBN-HOAS} and Section~\ref{sec-CBV}. 	
	 The relevant theories of this session are:
\begin{itemize}
	\item \textsf{CBN}, \textsf{Henkin},  \textsf{CBN$\_$CR}, \textsf{CBN$\_$Std} and \textsf{HOAS} for the CBN calculus
	\item \textsf{CBV}, \textsf{CBV$\_$CR} and \textsf{CBV$\_$Std} for the CBV calculus
\end{itemize} 

The theory \textsf{CBN} defines Section~\ref{sec-CBN}'s various reduction relations and proves basic facts about them, including fresh rule induction and fresh inversion principles. The relations have the following names in the formalization:
\begin{itemize}
\item The one-step reduction $\!\!\redS\!\!$ (Def.~\ref{def-CBN-red}) is \textsf{redn}. 
\item The one-step parallel reduction $\!\!\redP\!\!$ (Def.~\ref{def-redp-cbn}) is \textsf{rednP}. 
\item The labeled one-step parallel reduction $\!\!\redP_{\!\!\!\_}\!$ (Def.~\ref{defi-labeledPar}) is \textsf{rednPN}.   
\item The one-step left reduction $\!\!\redL\!\!$ (Def.~\ref{def-redL-cbn}) is \textsf{rednL}.  
\item The multi-step versions of the relations have an ``\textsf{M}'' prefixing their name: 
\textsf{Mredn}, \textsf{MrednP}, \textsf{MrednPN} and \textsf{MrednL}. 
\end{itemize}
Each of these relations also has infix notations. \textsf{redn}, \textsf{rednP}, \textsf{rednP} and \textsf{rednPN} are defined using Isabelle's \textsf{inductive} command, and their multi-step counterparts are defined by applying the reflexive-transitive closure operator from the Isabelle library.

The other mentioned theories have self-explanatory names: 
\begin{itemize}
	\item \textsf{Henkin} handles the soundness theorem for Henkin-style models (Section~\ref{sub-CBN-sound})
	\item \textsf{CBN$\_$CR} handles the Church-Rosser theorem (Section~\ref{sub-CBN-CR})
	\item \textsf{CBN$\_$Std} handles the standardization theorem (Section~\ref{sub-CBN-Std})
	\item \textsf{HOAS} handles the HOAS development (Section~\ref{sub-CBN-HOAS})
\end{itemize}

These theories also define the following recursive functions presented in this paper. In all cases, the end-product formal facts are obtained after expanding the definition of FSb or FSw model morphisms. 
\begin{itemize}
	\item Section~\ref{sub-inst-CBN}'s number of free occurrences operator,  $\no$, using substitution-aware recursion---Def.~\ref{no-def} corresponds to 
	\textsf{CBN}'s
	lemmas \textsf{no$\_$simps, no$\_$subst and no$\_$fresh}.
	\item Section~\ref{sub-CBN-CR}'s complete development operator, $\cdev$, using swapping-aware recursion---Def.~\ref{def-cdev} corresponds to theory \textsf{CBN$\_$CR}'s lemmas \textsf{``cdev$\_$simps 1'', 
		cdev$\_$App$\_$isDred,  
		cdev$\_$App$\_$isLm, 
		cdev$\_$App$\_$not$\_$isDred$\_$isLm and 
		 cdev$\_$swap and cdev$\_$fresh}. 
	 \item Section~\ref{sub-CBN-HOAS}'s HOAS encoding operator $\enc$---Def.~\ref{defi-hoas-enc} corresponds to theory \textsf{HOAS}'s lemmas 
	 \textsf{ enc$\_$simps, enc$\_$subst and enc$\_$fresh}.
\end{itemize} 

Finally, here is the mapping between main theorems presented in Section~\ref{sec-CBN} and their formal counterparts:
\begin{itemize} 
	 \item The Church-Rosser 
	 Theorem~\ref{th-CBN-CR} corresponds to theory \textsf{CBN$\_$CR}'s \textsf{theorem Mredn$\_$confluent}
	 \item The standardization  
	 Theorem~\ref{th-CBN-std} corresponds to theory \textsf{CBN$\_$Std}'s \textsf{theorem standardization}
	 \item The syntactic adequacy theorem represented by clauses (6)--(8) in Def.~\ref{defi-hoas-enc} 
	 corresponds to theory \textsf{HOAS}'s  
	 lemmas \textsf{enc$\_$subst, enc$\_$fresh and enc$\_$inj}.
   \item The $\beta$-reduction adequacy Theorem~\ref{thm-operAde} corresponds to theory \textsf{HOAS}'s \textsf{theorems 
	 	enc$\_$preserves$\_$rednL, 
	 	enc$\_$reflects$\_$MrednL 
	 	and rednL$\_$enc$\_$MrednL}.
\end{itemize}